\documentclass[aps]{revtex4}
\def\pochh #1#2{{(#1)\raise-4pt\hbox{$\scriptstyle#2$}}}
\def\binom#1#2{\left(\begin{array}{c}#1\\#2\end{array}\right)}
\begin {document}

\title {Functional integral with $\varphi^4$ term in the action \\ beyond
standard perturbative methods}
\author{J. Boh\' a\v cik}
\email{bohacik@savba.sk}
\affiliation{Institute of Physics, Slovak
Academy of Sciences, D\' ubravsk\' a cesta 9, 845 11 Bratislava,
Slovakia.}
\author{P. Pre\v snajder}
\email{presnajder@fmph.uniba.sk}
\affiliation{Department of
Theoretical Physics and Physics Education, Faculty of Mathematics,
Physics and Informatics, Comenius University, Mlynsk\' a dolina F2,
842 48 Bratislava, Slovakia.}

\begin{abstract}
We propose the another, in principle nonperturbative, method of the
evaluation of the Wiener functional integral for $\varphi^4$ term in
the action. All infinite summations in the results are proven to be
convergent. We find the "generalized" Gelfand-Yaglom differential
equation implying the functional integral in the continuum limit.
\end{abstract}
\maketitle

\section*{Introduction}

The conventional functional integral calculations rely on the
Gaussian method of the integration. This means, that physical
problems having the second order term in the action are calculable
precisely, those problems with higher orders terms in the action
must be calculated perturbatively. The perturbation methods are
applicable in plenty of the problems of the contemporary physics,
however, more frequently than ever recently we are confronted with
problems, where conventional perturbation methods are not
sufficient. The non-perturbative numerical methods are
successfully applied to answer the questions in the statistical
physics, quantum field theory, if we mentions only the well known
examples. The numerical methods give the answers "yes" or "no" on
the quantitative questions, the deal of the answer, "why" is still
missing.

We are going to discuss the analytical evaluation  of the simplest
case of the beyond Gaussian functional integral calculations,
where the action possesses  $\varphi^4$ term in the exponent. By
the convention, we call the term connected with $\varphi^4$ the
"coupling constant". In the conventional perturbation calculations
for $\varphi^4$ theory the results of the functional integral
calculations are obtained in the form of the power series of the
coupling constant (see, i.e., \cite{jin}). These series are
asymptotic, divergent, but by sophisticated re-summation
procedures one can obtain the reasonable results, allowing to take
the path integral more seriously.

In this article we propose the calculation  the Wiener functional
integral with term of the fourth order in the exponent by the
another method as in the conventional perturbation approach. In
contrast to the conventional perturbation theory, we expand into
the power series the term linear in the integration variable in
the exponent. In such case we can profit from the representation
of the functional integral by the parabolic cylinder functions. We
show, that in such case the expansions into the series are
uniformly convergent and we find the recurrence relations for the
Wiener functional integral in the $N$ - dimensional approximation.
We find the continuum limit of this finite dimensional integral by
procedure proposed by Gelfand-Yaglom \cite{bkz} for continuum
limit of the functional integral for the harmonic oscillator.

The article is organized as follows. In part 2 on the example of
the one dimensional integral we remember the calculation of the
integrals with the fourth order term in exponent by the parabolic
cylinder functions. In the part 3 we calculate the $N$ dimensional
integral as  the approximation of the functional integral. We will
prove the uniform convergence of the result. In the part 4 we will
show on the example of the "Independent Value Model" in the sense
introduced by Klauder \cite{klauder}, that this field theory model
we can solve non-perturbatively and we discuss the problem of
"triviality" of this model. In part 5 and in the appendices 1 and
2 we will study the result of the part 3. In the appendix 3 the
"generalized" Gelfand-Yaglom \cite{bkz} equation determining the
continuum limit of the functional integral is proven.  In the
appendix 4 we present the same result as in part 5 for
unconditional Wiener measure functional integral.

\section*{Simple integral}

We will define the functional integral as the limit of the finite
dimensional integral. This definition has not problems with the
idea of the integral measure, because for the finite dimensional
integrals the integral measure  is defined correctly. We explain
our method of calculation on the example of the one dimensional
integral. Calculating the finite dimensional integrals we must
solve the problem of the calculation of the integral
\begin {equation}
I_1=\int\limits _{-\infty}^{+\infty}\;dx\;\exp\{-(a x^4+b x^2+c
x)\}
 \label {int1d}
\end {equation}
where $Re\: a>0$.

\noindent We are interested also to the problems, when the fourth
order term is not small, and therefore it is not possible to treat
it as a perturbation comparing to the rest of the action term. The
generally accepted perturbative approach rely on the Taylor's
decomposition of the fourth order term with consecutive
replacements of the integration and summation order:
\begin {equation}
I_1=\sum\limits _{n=0} ^{\infty} \frac{(-a)^n} {n!}\int\limits
_{-\infty}^{+\infty}\;dx\;x^{4 n}\exp\{-(b x^2+c x)\}
\end {equation}
\noindent The integrals in above relation can be calculated, but
sum is
 divergent.

 However, $I_1=I_1(a,b,c)$ is an entire function for any complex values
 of $b$ and $c$, since there exist all integrals
 \[ \partial_c^n\partial_b^m I_1(a,b,c)=(-1)^{n+m}
 \int\limits _{-\infty}^{+\infty}
 \;dx\;x^{2m+n}\exp\{-(a x^4+bx^2+cx)\}
 \]
Consequently, the power expansions of $I_1=I_1(a,b,c)$ in $c$
and/or $b$ has an infinite radius of convergence (and in
particular they are uniformly convergent on any compact set of
values of $c$ and/or $b$). Let us now consider the power expansion
in $c$ which we shall frequently use:
\begin {equation}
I_1=\sum\limits _{n=0} ^{\infty} \frac{(-c)^n} {n!}\int\limits
_{-\infty}^{+\infty}\;dx\;x^n\exp\{-(a x^4+b x^2)\}
 \label{sim1}
\end {equation}
The integrals here appearing can be expressed in terms of the
parabolic cylinder function $D_{\nu}(z)$, $\nu=-m-1/2$, (see, for
instance, \cite {bateman}). For $n$ odd, due to symmetry of the
integrand the integrals are zero, for $n$ even, $n=2m$ we have:
 \[ D_{-m-1/2}(z)\ =\
\frac{e^{-z^2/4}}{\Gamma(m+1/2)}\int_0^\infty\;dx\;
 x^{m-1/2}\exp\{-\frac{1}{2}x^2-zx\} \]
 \begin {equation}
 \ =\ \frac{e^{-z^2/4}}{\Gamma(m+1/2)}\int_{-\infty}^{+\infty}\;dy\;
 y^{2m}\exp\{-\frac{1}{2}y^4-zy^2\}\ =\ \frac{(-1)^m}{\Gamma
 (m+1/2)}e^{-z^2/4}\partial^m_z\;e^{z^2/4}D_{-1/2}(z)\ .
\label{s1}
\end {equation}
\noindent Explicitly, for the Eq.(\ref{sim1}) we have:
 \begin {equation}
 I_1=\frac{\Gamma(1/2)}{(2a)^{1/4}}\sum\limits _{m=0} ^{\infty}
 \frac{(\xi)^m}{m!}e^{z^2/4}D_{-m-1/2}(z)
 \label{s2}\ ,\ \xi=\frac{c^2}{4\sqrt{2a}}\ ,\ z=\frac{b}{\sqrt{2a}}
\end {equation}
This sum is convergent for any values of $c$, $b$ and $a$
positive.

 The convergence of the infinite series (\ref{s2}) can be shown as follows.
 For $|z|$ finite, $|z|<\sqrt{|\nu|}$ and
$\mid arg(-\nu)\mid \leq \pi/2$, if $\mid \nu \mid \rightarrow
\infty$, the following asymptotic relation is valid\cite{bateman}:
\begin {equation}
D_{\nu}(z)=\frac{1}{\sqrt{2}}\; \exp\left[\frac{\nu}{2}
(\ln{(-\nu)}-1) -\sqrt{-\nu}\;
z\right]\left[1+O\left(\frac{1}{\sqrt{\mid \nu\mid
}}\right)\right] \label {ass1}
\end {equation}
The $m-th$ term  of the sum (\ref {s2}) possesses the asymptotic
\begin {equation}
\frac{1}{m!}\exp\left[-\frac{m}{2}(\ln{m}-1)
-\sqrt{m}\;z+z^2/4+m\ln\xi \right]
\end {equation}

 This means, following
Bolzano-Cauchy's criteria, that the sum (\ref {s2}) is not only
absolutely, but uniformly convergent also for the finite values of
the constants of the integral (\ref {int1d}).

\section*{The Calculation of the Functional Integral }

In this section we will calculate the functional integral defined
as the continuum limit of the finite dimensional integral. The
continuum limit means the limit $N\rightarrow\infty$, where $N$ is
the number of the time slices in the integral. The advantage of
this method rely in the well defined measure of the finite
dimensional integral. Taking the continuum limit of the result of
the $N$ dimensional integration, we bypass the problem of the
continuum integral measure.

We are going to calculate Wiener functional integral for the
continuum action defined by:
\begin {equation}
\mathcal{L} =\int \limits _0^\beta d\tau \left[c/2
\left(\frac{\partial\varphi(\tau)}{\partial\tau}\right)^2+b\varphi(\tau)^2
+a\varphi(\tau)^4\right]\ .
\end {equation}
Following the standard procedure, we divide the integration
interval to the $N$ equal slices, and we define the $N-th$
approximation of the continuum action:
$$\mathcal{L}_{N} = \sum\limits _{i=1}^N \triangle\left[c/2
\left(\frac{\varphi_i-\varphi_{i-1}}{\triangle}\right)^2
+b\varphi_i^2+a\varphi_i^4\right],$$ where $\triangle=\beta/N.$ By
definition of the integrals we have: $$
\mathcal{S}=\lim_{N\rightarrow \infty}\;\mathcal{S}_{N}\ .$$ We
define the $N-$ dimensional integral by the relation \cite{chai}:
\begin {equation}
\mathcal{Z}_{N}=\int\limits _{-\infty}^{+\infty} \prod \limits
_{i=1}^N\left(\frac{d\varphi_i}{\sqrt{\frac{2\pi\triangle}{c}}}\right)
\exp\left\{-\sum\limits _{i=1}^N \triangle\left[c/2
\left(\frac{\varphi_i-\varphi_{i-1}}{\triangle}\right)^2
+b\varphi_i^2+a\varphi_i^4\right]\right\}\ . \label {pcf3}
\end {equation}
The Wiener unconditional measure functional integral is
defined by formal limit:
$$\mathcal{Z} = \lim_{N\rightarrow \infty}\;\mathcal{Z}_{N}\ .$$
We use the formal notation for the continuum Wiener functional
integral:
$$\mathcal{Z} = \int [\mathcal{D}\varphi(x)]\exp \left\{-\int\limits_0^{\beta} d\tau \left[c/2
\left(\frac{\partial\varphi(\tau)}{\partial\tau}\right)^2+b\varphi(\tau)^2
+a\varphi(\tau)^4\right] \right\}\ .$$

The most important problem is to calculate the $N-$dimensional integral
(\ref{pcf3}). We rewrite the action in the convenient form for the
consecutive integrations:
\begin {eqnarray}
\mathcal{L}_N& = &\triangle a \varphi_1^4\ + \
\frac{c}{\triangle}(1+\frac{b \triangle^2}{c}) \varphi_1^2\ - \
\frac{c}{\triangle} \varphi_1 \varphi_2 \nonumber \\ & + & \cdots \nonumber\\
& + & \triangle a \varphi_i^4\ + \ \frac{c}{\triangle}(1+\frac{b
\triangle^2}{c}) \varphi_i^2\ - \ \frac{c}{\triangle} \varphi_i
\varphi_{i+1} \\ & + & \cdots \nonumber\\ & + & \triangle a
\varphi_N^4\ + \ \frac{c}{\triangle}(1/2+\frac{b \triangle^2}{c})
\varphi_N^2 \nonumber
\end {eqnarray}

In the integration over variable $\varphi_i$ one find the terms resulting from
the preceding integration over variable  $\varphi_{i-1}$, as one create the terms
for integration over variable $\varphi_{i+1}$. Exploating the formula \cite{prud}:

\begin {equation}
\int_0^{\infty}\ x^{\alpha-1}\ \exp (-p x^2-q x)\ d x = \Gamma (\alpha)(2 p)^{-\alpha/2} \exp\left(\frac{q^2}{8 p}\right)
D_{-\alpha}\left(\frac{q}{\sqrt{2p}}\right),
\label {prudnik1}
\end {equation}

for integration over the variable $\varphi_1$ first, we find:

\begin {equation}
\mathcal{Z}_1 =\int\limits _{-\infty}^{+\infty} \left(\frac{d\varphi_1}{\sqrt{\frac{2\pi\triangle}{c}}}\right)
\exp{\left\{-\triangle a \varphi_1^4\ - \
\frac{c}{\triangle}(1+\frac{b \triangle^2}{c}) \varphi_1^2\ + \
\frac{c}{\triangle} \varphi_1 \varphi_2\right\}}\ .
\end {equation}
By Taylor's expansions of the term $$\exp{\left(\frac{c}{\triangle} \varphi_1 \varphi_2\right)},$$
taking  into account that
terms of the odd powers in $\varphi_1$ disappears due to the symmetry of integral, we simplify the integral by
substitution: $$\varphi_1^2 = \omega .$$ We reads:

\begin {equation}
\mathcal{Z}_1 =\sum\limits _{k_1=0}^{\infty}\frac{(c \varphi_2 / \triangle)^{2 k_1}}{(2 k_1)!}
\int\limits _{0}^{+\infty} \left(\frac{d\omega}{\sqrt{\frac{2\pi\triangle}{c}}}\right)
(\omega)^{k_1-1/2}\exp{\left\{-\triangle a \omega^2\ - \
\frac{c}{\triangle}(1+\frac{b \triangle^2}{c}) \omega\right\}}\ .
\end {equation}
Using formula (\ref{prudnik1}) we find:

\begin {equation}
\mathcal{Z}_1 =\sum\limits _{k_1=0}^{\infty}\frac{(c \varphi_2 / \triangle)^{2 k_1}}{(2 k_1)!}
\left(\frac{1}{\sqrt{\frac{2\pi\triangle}{c}}}\right)
\Gamma(k_1+1/2)\left(\frac{1}{\sqrt{2 \triangle a}}\right)^{k_1+1/2}
\exp{(z^2/4)}\ D_{-k_1-1/2}(z)\ ,
\end {equation}
where $$z=\frac{c(1+b\triangle^2/c)}{\sqrt{2a\triangle^3}}\ .$$
Following definition of $z$, we can use the factor
$$\left(\frac{c}{\triangle}(1+b\triangle^2/c)\right)^{k_1+1/2}$$
to simplify further this relation to the form:

\begin {equation}
\mathcal{Z}_1 = \frac{1}{\sqrt{2\pi(1+b\triangle^2/c)}}\;
\sum\limits _{k_1=0}^{\infty}\frac{(\varphi_2)^{2 k_1}}{(2 k_1)!}
\left(\frac{c/\triangle}{(1+b\triangle^2/c)}\right)^{k_1}
\Gamma(k_1+1/2)\ z^{k_1+1/2}
\exp{(z^2/4)}\ D_{-k_1-1/2}(z)\ ,
\end {equation}

or, equivalently:

\begin {equation}
\mathcal{Z}_1  = \frac{1}{\sqrt{2\pi(1+b\triangle^2/c)}}\;
\sum\limits _{k_1=0}^{\infty}\; \frac{\left(\frac{\scriptstyle
c}{\scriptstyle
\triangle(1+b\triangle^2/c)}\right)^{k_1}}{(2k_1)!}\;(\varphi_2)^{2k_1}\;
\Gamma(k_1+1/2)\mathcal{D}_{-k_1-1/2}(z)\ , \nonumber
\end {equation}
where the notation was used:
\begin {equation}
 \mathcal{D}_{-k_1-1/2}(z) = z^{k_1+1/2}
e^{\frac{\scriptstyle z^2}{\scriptstyle 4}}\; D_{-k_1-1/2}(z)\ .
\end {equation}
From above result, the term $(\varphi_2)^{2k_1}$ will be involved
into the second integration over the variable $\varphi_2$. This
will results to the additional $k_1$ dependent contribution to the
result of the first integration and to the introduction of the link
between both steps of integration. Explicitly, for the second
integration, we have:
\begin {equation}
\mathcal{Z}_2^{loc} = \int\limits
_{-\infty}^{+\infty}\;\frac{d\varphi_2}{\sqrt{\frac{2\pi\triangle}{c}}}\;
(\varphi_2)^{2k_1}\;\exp\left\{-\triangle a \varphi_2^4\ - \
\frac{c}{\triangle}(1+\frac{b \triangle^2}{c}) \varphi_2^2 +
\frac{c}{\triangle} \varphi_2 \varphi_3 \right\}\ .
\end {equation}
By Taylor's expansion of the term of the action linear in variable
$\varphi_2$, by the interchange of the order of the integration
and summation, taking into account that the odd powers of $\varphi_2$ variable disappear due to symmetry of integral,
and by variable substitution $\varphi^2_2 = \omega$  we find:

\begin {equation}
\mathcal{Z}_2^{loc} =\sum\limits _{k_2=0}^{\infty}\frac{(c \varphi_3 / \triangle)^{2 k_2}}{(2 k_2)!}
\int\limits _{0}^{+\infty} \left(\frac{d\omega}{\sqrt{\frac{2\pi\triangle}{c}}}\right)
(\omega)^{k_2+k_1-1/2}\exp{\left\{-\triangle a \omega^2\ - \
\frac{c}{\triangle}(1+\frac{b \triangle^2}{c}) \omega\right\}}\ .
\end {equation}
By the same evaluation as for $Z_1$ we have:

\begin {equation}
\mathcal{Z}_2^{loc} = \frac{\left(c(1+b\triangle^2/c)/\triangle\right)^{-k_1}}{\sqrt{2\pi(1+b\triangle^2/c)}}\;
\sum\limits _{k_2=0}^{\infty}\; \frac{\left(\frac{\scriptstyle
c}{\scriptstyle
\triangle(1+b\triangle^2/c)}\right)^{k_2}}{(2k_2)!}\;(\varphi_3)^{2k_2}\;
\Gamma(k_2+k_1+1/2)\mathcal{D}_{-k_2-k_1-1/2}(z)\ . \nonumber
\end {equation}

Taking both integration steeps together we have:
\begin {equation}
\mathcal{Z}_2 =
\left(\frac{1}{\sqrt{2\pi(1+b\triangle^2/c)}}\right)^2  \sum
\limits_{k_1,k_2=0}^\infty  \frac
     {\xi^{2k_1} \left(\frac{\scriptstyle
c}{\scriptstyle
\triangle(1+b\triangle^2/c)}\varphi_3^2\right)^{k_2}}
     {(2k_1)!(2k_2)!}\;
\Gamma(k_1+1/2)\mathcal{D}_{-k_1-1/2}(z)\;\Gamma(k_1+k_2+1/2)\mathcal{D}_{-k_1-k_2-1/2}(z)\ ,
\end {equation}
\noindent where the new symbol was introduced:
$$\xi = \frac{1}{(1+b\triangle^2/c)}\ .$$

We can repeats this procedure for the integration variables $\varphi_3, \cdots, \varphi_{N-1}.$ For the integration over
variable $\varphi_N$, there one have no linear term in exponent, therefore we don't expand anything and this last integration
don't add the summation to the final formula. We have:

\begin {equation}
\mathcal{Z}_N^{loc} = \int\limits
_{-\infty}^{+\infty}\;\frac{d\varphi_N}{\sqrt{\frac{2\pi\triangle}{c}}}\;
(\varphi_N)^{2k_{N-1}}\;\exp\left\{-\triangle a \varphi_N^4\ - \
\frac{c}{\triangle}(1/2+\frac{b \triangle^2}{c}) \varphi_N^2 \right\}\ .
\end {equation}
and the result is:

\begin {equation}
\mathcal{Z}_N^{loc} =
\frac{\left(c(1/2+b\triangle^2/c)/\triangle\right)^{k_{N-1}}}{\sqrt{2\pi(1/2+b\triangle^2/c)}}\;
\Gamma(k_{N-1}+1/2)\mathcal{D}_{-k_{N-1}-1/2}(z_{N})\ . \nonumber
\end {equation}

Repeating by recurrence this procedure, we obtain finally the
non-perturbative, exact  result:
\begin {equation}
\mathcal{Z}_{N} =
\left[2\pi(1+b\triangle^2/c)\right]^{-\frac{N-1}{2}}
\left[2\pi(1/2+b\triangle^2/c)\right]^{-1/2}
\sum\limits_{k_1,\cdots,k_{N-1}=0}^\infty \prod \limits _{i=1}^N \;
\left[ \frac{\left(\xi_i\right)^{2k_{i}}}{(2k_{i})!}
\Gamma(k_{i-1}+k_{i}+1/2)\mathcal{D}_{-k_{i-1}-k_{i}-1/2}\;(z_i)\right]\ ,
\label{fin1}
\end {equation}
\noindent where $k_0 \equiv k_N \equiv 0,$ and $\xi_1 = \xi_2 =
\cdots = \xi_{N-2} = \xi,\ \xi_N=0$ also $z_1 = z_2 = \cdots =
z_{N-1} = z.$ The modifications appears due to the last integration
step and we have $$z_{N} =
\frac{c(1/2+b\triangle^2/c)}{\sqrt{2a\triangle^3}}\ , \; \xi_{N-1} =
\sqrt{\frac{1}{(1+b\triangle^2/c)}}\
\sqrt{\frac{1}{(1/2+b\triangle^2/c)}}\ .
$$
For the conditional measure functional integral, the variable $\varphi_N$ will be fixed,
in time-slicing procedure the last integral over variable $\varphi_N$ is absent. This means,
that we'll have the same values for all $\xi_i$ as well as for $z_i$, and upper limit for product formula
in (\ref{fin1}) should be $N-1$.

In the noninteracting harmonic oscillator limit, ($b$ positive,
$a=0$), the Eq.(\ref{fin1}) is reduced to the $N$-dimensional
approximation to the unconditional measure Wiener functional
integral for the harmonic oscillator. For $a\neq 0$ we can derive
from  (\ref{fin1}) by Ge\`lfand-Yaglom procedure the differential
equation of the second order. The dependence on the coupling
constant is in the part characterized by multiple summations over
indices $k_i$. This part is now the object of our discussion.

We are going to show that relation (\ref{fin1}) for the
$N-$dimensional integral is uniformly convergent for each of the
summation index. For the summation over one index $k_i$ we have (we omit the index $i$
in $\xi_i$ and $z_i$):
\begin {equation}
\sum \limits_{k_i=0}^\infty
\left[\frac{\left(\xi\right)^{2k_{i}}}{(2k_{i})!}
\Gamma(k_{i-1}+k_{i}+1/2)\mathcal{D}_{-k_{i-1}-k_{i}-1/2}\;(z)\right]
\left[
\Gamma(k_{i}+k_{i+1}+1/2)\mathcal{D}_{-k_{i}-k_{i+1}-1/2}\;(z)\right]\ .
\label{sum1}
\end {equation}
For $k_i \rightarrow \infty$ we apply for the $D$ functions the asymptotic relation
(\ref{ass1}). We have for the $k_i$ dependent part of
(\ref{sum1}) the relation:
\begin {eqnarray}
a_{k_i} &\equiv& \frac{(\xi)^{2k_{i}}\Gamma(k_{i}+k_{i+1}+1/2)\Gamma(k_{i-1}+k_{i}+1/2)}
{(2k_{i})!}\; \exp{(z^2/2)}
\label{defi1}\\
\noalign{\vskip8pt} &\times& z^{k_{i}+k_{i+1}+1/2}\;
\exp\left[-\frac{k_{i}+k_{i+1}}{2}(\ln{(k_{i}+k_{i+1})}-1)
-\sqrt{k_{i}+k_{i+1}}\; z\right]\nonumber \\
\noalign{\vskip8pt} &\times& z^{k_{i-1}+k_{i}+1/2}\;
\exp\left[-\frac{k_{i-1}+k_{i}}{2}(\ln{(k_{i-1}+k_{i})}-1)
-\sqrt{k_{i-1}+k_{i}}\; z\right]\nonumber
\end {eqnarray}
We convert the above relation by help of the
identity:
$$\Gamma(p+k+1/2)=\exp{[\ln{\Gamma(p+k+1/2)}]}$$
to the form convenient for the proof of the convergence.

For the logarithm of the relation (\ref{defi1}) we have:
\begin {eqnarray}
\ln{a_{k_i}} &=& \ln\Gamma(k_{i}+k_{i+1}+1/2)+
\ln\Gamma(k_{i-1}+k_{i}+1/2)
-\ln\Gamma(2k_{i}+1)\label{ki1}\\
\noalign{\vskip8pt}
&-&\frac{k_{i-1}+k_{i}}{2}(\ln{(k_{i-1}+k_{i})}-1)
-\frac{k_{i}+k_{i+1}}{2}(\ln{(k_{i}+k_{i+1})}-1) +2k_i\ln (\xi z)
+(k_{i-1}+k_{i+1}+1)\ln z \nonumber  \\
&-&(\sqrt{k_{i}+k_{i+1}}+\sqrt{k_{i}+k_{i-1}})z + z^2/2
+o(k_i^{-1/2}) \nonumber
\end {eqnarray}
For  the gamma functions with argument $u \rightarrow\infty $ we
can use the asymptotic relation \cite{bateman1}:
\begin {equation}
\ln\Gamma(u) = (u-1/2)\ln(u)-u+1/2\ln(2\pi)+\mathcal{O}(u^{-1})
\label{stirlitz}
\end {equation}

For $k_{i-1}$ and $k_{i+1}$ fixed and finite and if $k_{i\pm 1}<k_i$, we find the following
asymptotic for $a_{k_{i}}$:
\begin {eqnarray}
\ln a_{k_i}&=& -(k_i - \frac{k_{i-1}+k_{i+1}-1}{2})\ln k_i +
2k_i(\ln(\xi z) -\ln2 +1/2)-2\sqrt{k_i}\; z + (k_{i-1}+k_{i+1}+1)\ln z
\nonumber \\
&+& 1/2\ln{\pi} + z^2/2 + o(k_i^{-1/2})
\end {eqnarray}
The leading term of the above relation is $$-k_i \ln(k_i)\ ,$$
and  $a_{k_i}$ is going to zero in the asymptotic region as
$$\frac{k_i^{\alpha}\beta^{k_i}}{k_i!\; \exp{(\sqrt{k_i})}}\; ,$$
where $\alpha$ and $\beta$ are finite numbers. The asymptotic
feature of $a_{k_i}$ is sufficient for proof of the uniform
convergence of the series (\ref{sum1}) and therefore for the
finiteness of the sum (\ref{sum1}).

By the same method we can prove the uniform convergence of the
summation over two, three, ..., $N-1$ summation indices $k_i$ in
the equation (\ref{fin1}). This important conclusion indicate,
that the $N$ dimensional approach to the the continuum functional
integral can be summed up.

\section*{Independent Value Model}

The another example of the use of the calculation of the functional
integral by help of the parabolic cylinder function is the problem
of the independent value model\cite{klauder}. The independent value
models (ivm) are covariant models without gradient terms. The formal
functional integral for generating functional of such models reads:

\begin {equation}
Z^{ivm}[J]=\mathcal{N}\int\; \mathcal{D}\phi\; \exp{\left\{\int
d^nx[i\;J(x)\phi(x)-\frac{1}{2}m^2\phi(x)^2-g\phi(x)^4]\right\}}\; ,
\label{ivm1}
\end {equation}
where $\phi(x)$ is a real field, $J(x)$ is the source term and $g$
is the interaction constant. The normalization constant
$\mathcal{N}$ guarantees that $Z^{ivm}[0]=1$. It should be noted,
that physically the model is trivial (the field equation reduces to
$\phi(x)=0$), corresponding to independent fluctuations for
different $x$'s. However, the functional integrals of this type have
a well-defined mathematical meaning, and its investigations can
serve as a tool for a better understanding of various approximative
schemes of calculations.

The generating functional $Z^{ivm}[J]$ of the ivm can be calculated
as the continuum limit of the discretized finite dimensional
integral approximation to the continuum functional integral
\begin {eqnarray}
 \prod\limits_{k=1}^N\; Z^{ivm}_k[J]
= \mathcal{N}_0 \prod\limits_{k=1}^N\; \int d\phi_k\;
\exp{\left\{\sum\limits_{k=1}^N(iJ_k\phi_k\triangle-1/2
m^2_0\phi^2_k\triangle - g_0\phi^4_k\triangle)\right\}}\;
\label{ivm2}  ,
\end {eqnarray}
where $\triangle$ is the volume of the discretized space cell, and
$\mathcal{N}_0$ is the normalization constant. The values $m_0$ and
$g_0$ are supposed to be both positive and $\triangle$ dependent. In
this model the $\triangle$ dependence is chosen so that in the
continuum limit $\triangle\rightarrow 0$ the physical observables in
the model take reasonable values. Only when this dependence is fixed
the model is defined. The $\ln Z[J]^{ivm}$ is the generating
functional for the connected correlation functions and all even
order connected correlations functions \emph{are nonnegative}, as
follows from the standard canonical form of the generating
functional for the ivm model. Performing a discretized calculation
Lebowitz \cite{lebo1} proved that the connected correlation
functions of the fourth order \emph{are non-positive}, when a free
measure of the functional integral was used. The two inequalities
for the connected four point correlation function are simultaneously
valid only if such function vanish. This means, that the discretized
approach to the functional integral leads to the gaussian generating
functional of a free, noninteracting model, with no dependence on
the coupling constant in the continuum limit. Such behavior is known
as "triviality".

We show that we can obtain a nontrivial result for the generating
functional of  ivm model {\it non-perturbatively} by taking a
particular $N\rightarrow\infty$ limit in (\ref{ivm2}) -- the
continuum limit of the discretized approximation. In evaluations of
finite dimensional integrals we follow the parabolic cylinder
functions representation \cite{bateman} of the integrals with $x^4$
in the exponent of the integrand:

$$ \int\limits _{-\infty}^{+\infty}\;x^{2m} dx\;\exp\{-(a
x^4+b x^2)\}\ \hskip4cm $$ 
\begin {equation} =\frac{1}{\left(2a\right)^{m/2+1/4}}\Gamma(m+1/2)\;
\exp{(\frac{b^2}{8a})}\;
D_{-m-1/2}\left(\frac{b}{\sqrt{2a}}\right)\; ,
\label{int1d1}
\end {equation}
where $Re\: a>0$, and $D_{\nu}(z)$ is the parabolic cylinder
function of index $\nu$ and argument $z$.

\noindent Applying (\ref{int1d1})  to the finite dimensional
approximation of the functional integral (\ref{ivm2}), we find for
the $k-th$ factor in (\ref{ivm2}) the result:
\begin {equation}
Z_k^{ivm}[J] = (\mathcal{N}_0)^{1/N} \sum\limits_{i=0}^{\infty}
\frac{\left(-\rho_k\right)^i}{i!}\; \left(z\right)^{i+1/2} \;
\exp{(\frac{z^2}{4})}D_{-i-1/2}\left(z\right)\; , \label{sul1}
\end {equation}
where $\rho_k=(J_k^2\triangle)/(2m^2_0)$ and
$z=(m_0^2\sqrt{\triangle})/\sqrt{8g_0}$. The sum in Eq.(\ref{sul1})
is uniformly convergent with respect to $z$ on any compact domain as
follows from the asymptotic expression for parabolic cylinder
functions when the index $|\nu|\rightarrow\infty$.

The definition of the discretized approach is accomplished when the
dependence of "bare" parameters in (\ref{ivm2}) on $\triangle$ (or
equivalently on $N$) is fixed in such way that continuum limit
$N\to\infty$ is well-defined. The requirement
$\lim_{\triangle\rightarrow 0} (g_0/\triangle)\neq 0$ (i.e.,
$z\rightarrow const$ in the continuum limit) is the necessary
condition for a finite continuum limit of the discretized ivm model.
Using (\ref{sul1}) the exact formula for the generating functional
(\ref{ivm2}) reads

\begin {equation}
Z^{ivm}[J]= \lim_{N\rightarrow \infty}\left( \prod_{k=1}^N
\left\{1 -\rho_k z \frac{D_{-3/2}(z)}{D_{-1/2}(z)} + \frac{1}{2}
\rho_k^2 z^2 \frac{D_{-5/2}(z)}{D_{-1/2}(z)} -
...\right\}\right)\; ,\label{fin1i}
\end {equation}
where we have taken into account the proper normalization
$Z^{ivm}[0]=1$. In the continuum limit $\sum \triangle$ is replaced
by $\int dx^n$. We see immediately that the continuum limit survive
only terms with the number of the summations equal to the power of
$\triangle$. This restricts the contributions to terms up to the
first order in $\rho_k$. In the limit $N\rightarrow \infty$ (i.e.,
$\triangle\rightarrow 0$) we obtain the nontrivial contribution to
the generating functional (\ref{ivm1}) when $m_0$ and $z^{-2}\sim
g_0 N$ are kept fixed:
\begin {equation}
Z^{ivm}[J]\ =\ \exp{\left\{- \kappa(z)\int
d^nx\frac{J^2(x)}{2m_0^2}\right\}}\; ,\
\kappa(z)=1-\frac{3D_{-5/2}(z)}{2D_{-1/2}(z)}\; ,\label{fin2}
\end {equation}
The result (\ref{fin2}) represents a free theory result, due to the
quadratical  $J^2$ dependence in the exponent. However, (\ref{fin2})
possesses a nontrivial dependence on interaction term which survives
in the continuum limit and is hidden in $z\sim
\sqrt{(\triangle/g_0)}$ dependence of the factor $\kappa(z)$
multiplying the integral in the exponent. Introducing a new
parameter $m^2(z,m_0^2)=\frac{1}{ \kappa(z)}m^2_0$ all non-trivial
dependence on $z$ and $m_0^2$ is included into mass renormalization.
This is a new feature, because within perturbative approach was
found no dependence on the interaction constant in the continuum
limit.

\section*{Summation over $k_i$}

The result of the $N$-dimensional integration (\ref{fin1}) is the
exact one, calculated without any approximation. The multiple
summations suppress this advantage somewhat, therefore we shall
attempt to provide the $k_i$ summations in the formula (\ref{fin1}).
The details of this calculations are in the Appendices 1 and 2, here
we sketch the method and results only.

To obtain the formula for subsequent evaluation of the functional
integral, we must to solve the problem of summations over indexes
$k_i$ in Eq. (\ref{fin1}). For each summation index $k_i$ we meet
the sum of the series with product of two $\mathcal{D}$ functions.
We are dealing with the sum of the series:

\begin {equation}
a_{k_i}=\frac{\left(\xi\right)^{2k_{i}}}{(2k_{i})!}
\Gamma(k_{i-1}+k_{i}+1/2)\mathcal{D}_{-k_{i-1}-k_{i}-1/2}\;(z)
\Gamma(k_{i}+k_{i+1}+1/2)\mathcal{D}_{-k_{i}-k_{i+1}-1/2}\;(z)
\label{ass}
\end {equation}

Of course, parabolic cylinder functions belong to the representation
of the upper-triangle matrices, therefore we believe to the
simplification of the product of two such functions. What follows,
we propose the our approach to provide the summation over indices
$k_i$. At first, we represent one of $\mathcal{D}$ functions by
Poincar\' e - type expansion \cite{bateman}, valid for real index
and positive argument of the function, if the inequality  $a < z$ is
valid:

\begin {equation}
e^{z^2/4}\;z^{a+1/2}\;D_{-a-1/2}(z)\;=\;
\mathcal{D}_{-a-1/2}(z)\;=\; \sum\limits_{j=0}^{\mathcal{J}}\;
(-1)^j \;\frac{(a + 1/2)_{2j}}{j!\;(2z^2)^j} +
\epsilon_{\mathcal{J}}(a,z)\ , \label{asex1}
\end {equation}
where $\epsilon_{\mathcal{J}}(a,z)$ is the remainder of the
Poincar\' e - type expansion of the $\mathcal{D}$ function.

Inserting this relation to the sum,  we divide the sum over $k_i$
into two parts. One, over finite $k_i$, where Poicar\'e - type
expansion is correct and the second, the remainder over infinite
$k_i$ indexes, but small comparing to first part due to uniform
convergence:
\begin {eqnarray}
&&\sum\limits_{j=0}^{\mathcal{J}}\; \frac{(-1)^j }{j!\;(2z^2)^j}\
\sum_{k_i=0}^{N_0}\ \frac{\left(\xi\right)^{2k_{i}}}{(2k_{i})!}
\Gamma(k_{i-1}+k_{i}+1/2) \Gamma(k_{i}+k_{i+1}+2j+1/2)
\mathcal{D}_{-k_{i-1}-k_{i}-1/2}\;(z)
  +\\
 &+&\sum_{k_i=0}^{N_0}\ \frac{\left(\xi\right)^{2k_{i}}}{(2k_{i})!}
\Gamma(k_{i-1}+k_{i}+1/2) \Gamma(k_{i}+k_{i+1}+1/2)
\mathcal{D}_{-k_{i-1}-k_{i}-1/2}\;(z) \
\epsilon_{\mathcal{J}}(a,z) + \nonumber \\
&+&\sum_{k_i=N_0+1}^{\infty}\
\frac{\left(\xi\right)^{2k_{i}}}{(2k_{i})!}
\Gamma(k_{i-1}+k_{i}+1/2) \Gamma(k_{i}+k_{i+1}+1/2)
\mathcal{D}_{-k_{i-1}-k_{i}-1/2}\;(z)\mathcal{D}_{-k_{i}-k_{i+1}-1/2}\;(z),\nonumber
\end {eqnarray}
where $N_0<z$. In the first term of the above relation we add the
terms expanding the summation over index $k_i$ to infinity and in
the asymptotic region of $ k_i>N_0$ we expand the function
$\mathcal{D}$ by double asymptotic properties expansions proposed by
Temme \cite{temme}:
\begin {equation}
\mathcal{D}_{-a-1/2}(z)=\frac{\exp{(-\mathcal{A}z^2)}}{(1+4\lambda)^{1/4}}
\left[\sum_{k=0}^{n-1}\frac{f_k(\lambda)}{z^{2k}}\ +\
\frac{1}{z^{2n}}R_n(a,z)\right]
\end {equation}
where the following quantities were introduced:
 $$\lambda = \frac{a}{z^2}\ ,\
 w_0 = \frac{1}{2}\; [\sqrt{1+4\lambda}-1]\ ,\
 \mathcal{A} =\frac{1}{2}\; w_0^2 + w_0 - \lambda - \lambda
 \ln{w_0} + \lambda\ln {\lambda}\ .$$
 The functions $f_k(\lambda)$ are calculated in \cite{temme}. For
 sum of the series (\ref{ass}) we find:

\begin {eqnarray}
&&\sum\limits_{j=0}^{\mathcal{J}}\; \frac{(-1)^j }{j!\;(2z^2)^j}\
\sum_{k_i=0}^{\infty}\ \frac{\left(\xi\right)^{2k_{i}}}{(2k_{i})!}
\Gamma(k_{i-1}+k_{i}+1/2) \Gamma(k_{i}+k_{i+1}+2j+1/2)
\mathcal{D}_{-k_{i-1}-k_{i}-1/2}\;(z)
  +\\
 &+&\sum_{k_i=0}^{N_0}\ \frac{\left(\xi\right)^{2k_{i}}}{(2k_{i})!}
\Gamma(k_{i-1}+k_{i}+1/2) \Gamma(k_{i}+k_{i+1}+1/2)
\mathcal{D}_{-k_{i-1}-k_{i}-1/2}\;(z) \
\epsilon_{\mathcal{J}}(a,z) + \nonumber \\
&+&\sum_{k_i=N_0+1}^{\infty}\
\frac{\left(\xi\right)^{2k_{i}}}{(2k_{i})!}
\Gamma(k_{i-1}+k_{i}+1/2) \Gamma(k_{i}+k_{i+1}+1/2)
\mathcal{D}_{-k_{i-1}-k_{i}-1/2}\;(z)\frac{\exp{(-\mathcal{A}z^2)}}{(1+4\lambda)^{1/4}}
\left[\sum_{k=0}^{n-1}\frac{f_k(\lambda)}{z^{2k}}\right]+\nonumber\\
&+&\sum_{k_i=N_0+1}^{\infty}\
\frac{\left(\xi\right)^{2k_{i}}}{(2k_{i})!}
\Gamma(k_{i-1}+k_{i}+1/2) \Gamma(k_{i}+k_{i+1}+1/2)
\mathcal{D}_{-k_{i-1}-k_{i}-1/2}\;(z)\frac{\exp{(-\mathcal{A}z^2)}}{(1+4\lambda)^{1/4}}
\left[
\frac{1}{z^{2n}}R_n(a,z)\right]-\nonumber\\
&-&\sum\limits_{j=0}^{\mathcal{J}}\; \frac{(-1)^j }{j!\;(2z^2)^j}\
\sum_{k_i=N_0+1}^{\infty}\
\frac{\left(\xi\right)^{2k_{i}}}{(2k_{i})!}
\Gamma(k_{i-1}+k_{i}+1/2) \Gamma(k_{i}+k_{i+1}+2j+1/2)
\mathcal{D}_{-k_{i-1}-k_{i}-1/2}\;(z) \label{rem}
\end {eqnarray}

 The sum over index $k_i$ in the first term of the above relation will be
provided by help of the  formula \cite{bateman}:
\begin {equation}
e^{x^2/4}\sum\limits_{k=0}^{\infty}\;
\frac{\pochh{\nu}k}{k!}\;t^k\;D_{-\nu-k}(x)\;=\;e^{(x-t)^2/4}\;D_{-\nu}\;(x-t)\
. \label{dsum1}
\end {equation}
Method of this sum is described in Appendix 1, we are going to
evaluate the three terms of remainder now. First of all, we must to
develop the criteria for the remainder. Schematically, we must
evaluate the multiple sums:
\begin {equation}
\sum^{\infty}_{k_1,\cdots ,k_N=0}a_{k_1, k_2}a_{k_2, k_3}\cdots
a_{k_{N-1}, k_N}a_{k_N, k_{N+1}}
\end {equation}
If each individual sum exist and it is finite, we divide the sum
over index $k_i$ to finite, principal sum and the remainder, which
can be done as small as possible, comparing to the principal part of
the sum. Let us define the principal sum object, where sum over $m$
indexes $k_i$ was not provided, but for indexes $k_{m+1,\cdots ,
k_N}$ the finite sum was done:
\begin {equation}
\Sigma_{m,N-m}=\sum^{\infty}_{k_1,\cdots ,k_m=0}a_{k_1, k_2}a_{k_2,
k_3} \cdots a_{k_{m-1}, k_m}b_{k_m},
\end {equation}
where
\begin {equation}
b_{k_m} = \sum^{N_0}_{k_{m+1},\cdots ,k_N=0}a_{k_{m},
k_{m+1}}a_{k_{m+1}, k_{m+2}} \cdots a_{k_{N}, k_{N+1}}
\end {equation}
The next sum over $k_{m}$ we divide into the finite, principal sum
and the remainder, the infinite sum, giving minor contribution. We
 have the identity:
\begin {equation}
\Sigma_{m,M-m}=\Sigma_{m-1,M-m+1}+\Sigma_{m-1,M-m}\ \epsilon_m
\end {equation}
where $$\epsilon_m = \sum_{k_m=N_0+1}^{\infty}a_{k_{m-1},
k_m}b_{k_m}.$$ Let $\epsilon=\max{(\epsilon_1, \epsilon_2, \dots ,
\epsilon_N)}$, then we find:
\begin {equation}
\Sigma_{N,0}\leq \Sigma_{0,N}+\binom{N}{1}\Sigma_{0,N-1}\
\epsilon+\cdots+\binom{N}{i}\Sigma_{0,N-i}\ \epsilon^i
+\cdots+\epsilon^N
\end {equation}
If $\Sigma_{0,k}$ are the members of non-decreasing series, the
following inequality is valid:
\begin {equation}
\Sigma_{0,N}\leq \Sigma_{N,0}\leq \Sigma_{0,N}(1+\epsilon)^N
\end {equation}
In the limit $N\rightarrow \infty$ we obtain
$\Sigma_{0,\infty}=\Sigma_{\infty,0}$, if $\epsilon\simeq
N^{-1-\varepsilon}$ where $\varepsilon>0.$ Let us check if the terms
in the remainder corresponds to this demand.

For Poincar\' e - type expansion the upper bound of remainder was
calculated by Olver \cite{olver}. We use the improved upper bound
evaluated by Temme \cite{temme2}. The upper bound for remainder in
definition (\ref{asex1}) can be read:
\begin {equation}
\mid \epsilon_{\mathcal{J}}(a,z)\mid\ \leq
\frac{2z^2}{z^2-2a}\frac{\pochh{a+1/2}{2\mathcal{J}}}{(\mathcal{J}-1)!\
(2z^2)^{\mathcal{J}}}\ _1F_2(\mathcal{J}/2, 1/2; \mathcal{J}/2+1;
1-\frac{a^2}{z^2})\exp{\left(\frac{4\delta}{z^2-2a}\
_1F_2(1/2,1/2;3/2;1-\frac{a^2}{z^2})\right)}
\end {equation}
This relation is valid for $2\sqrt{a}\leq z$. The following quantity
was introduced:
$$\delta=\Big|\frac{a^2}{4}+\frac{3}{16}\Big|+
\frac{2a}{z^2}\left(1+\frac{a}{2z^2}\right)\frac{z^2}{(z^2-2a)^2}$$

The sum in remainder:
\begin {equation}
\sum_{k_i=0}^{N_0}\ \frac{\left(\xi\right)^{2k_{i}}}{(2k_{i})!}
\Gamma(k_{i-1}+k_{i}+1/2) \Gamma(k_{i}+k_{i+1}+1/2)
\mathcal{D}_{-k_{i-1}-k_{i}-1/2}\;(z) \
\epsilon_{\mathcal{J}}(k_i+k_{i+1},z)
\end {equation}
is bounded by the relation:
\begin {equation}
\frac{\mathcal{M}}{(\mathcal{J}-1)!\
(2z^2)^{\mathcal{J}}}\sum_{k_i=0}^{N_0}\
\frac{\left(\xi\right)^{2k_{i}}}{(2k_{i})!}
\Gamma(k_{i-1}+k_{i}+1/2) \Gamma(k_{i}+k_{i+1}+2\mathcal{J}+1/2)
\mathcal{D}_{-k_{i-1}-k_{i}-1/2}\;(z) \
\end {equation}
where
\begin {equation}
\mathcal{M}=\max{\left(\frac{2z^2}{z^2-2a} \cdot_1F_2(\mathcal{J}/2,
1/2; \mathcal{J}/2+1;
1-\frac{a^2}{z^2})\exp{\left(\frac{4\delta}{z^2-2a}\
_1F_2(1/2,1/2;3/2;1-\frac{a^2}{z^2})\right)}\right)\ ,}
\end {equation}
and $a=k_i+k_{i+1}$, $k_i=1, 2, \dots, N_0\ .$ We have the freedom
to choose the parameter $\mathcal{J},$ therefore the upper bound on
this contribution to the remainder can be done as small as
necessary.

For the remainder in the double asymptotic property expansion
\cite{temme} we find the definition \cite{temme2}:
\begin {equation}
R_n(a,z)=\frac{z^{2a+1/2}}{\Gamma(2a+1/2)}\int_0^{\infty}s^a e^{-z^2
s}\ f_n(s)\frac{d s}{\sqrt{s}}
\end {equation}
It can be shown by some algebra, that upper bound on remainder is:
$$R_n(a,z)\leq M_n\ ,$$ where $M_n$ is upper bound to the function
$f_n(s)$.

We are now going to estimate the upper bound of the following
contribution to the remainder in (\ref{rem}):
\begin {equation}
\sum_{k_i=N_0+1}^{\infty}\
\frac{\left(\xi\right)^{2k_{i}}}{(2k_{i})!}
\Gamma(k_{i-1}+k_{i}+1/2) \Gamma(k_{i}+k_{i+1}+1/2)
\mathcal{D}_{-k_{i-1}-k_{i}-1/2}\;(z)\frac{\exp{(-\mathcal{A}z^2)}}{(1+4\lambda)^{1/4}}
\left[ \frac{1}{z^{2n}}R_n(a,z)\right]_{a=k_i+k_{i+1}} \label {rem3}
\end {equation}
In the asymptotic region of $k_i$ we use the Stirling formula for
the gamma functions:
\begin {equation}
\ln{{\Gamma(u)}}=(u-1/2)\ln u - u + 1/2\ln{(2 \pi)} +
\mathcal{O}(u^{-1})
\end {equation}
We find:
\begin {equation}
\frac{\left(\xi\right)^{2k_{i}}}{(2k_{i})!}
\Gamma(k_{i-1}+k_{i}+1/2)
\Gamma(k_{i}+k_{i+1}+1/2)=\left(\frac{\xi}{2}\right)^{2k_{i}}\left(k_i\right)^{k_{i+1}+k_{i-1}-1/2}
\exp{\left(\frac{k_{i+1}(k_{i+1}+1/2)+k_{i-1}(k_{i-1}+1/2)}{k_i}\right)}
\label{eg21}
\end {equation}
Moreover, in the asymptotic region of $k_i$ we replace the function
$\mathcal{D}_{-k_{i-1}-k_{i}-1/2}\;(z)$ by the principal part of the
double asymptotic expansion:
\begin {equation}
\mathcal{D}_{-k_{i-1}-k_{i}-1/2}\;(z)=\frac{\exp{(-\mathcal{A}z^2)}}{(1+4\lambda)^{1/4}}\Big
|_{\lambda=\frac{k_{i-1}+k_{i}}{z^2}}
\end {equation}
Taking into account that
\begin {equation}
-\mathcal{A}z^2 = -\frac{\lambda
z^2}{2}\left(-1+\frac{2}{1+\sqrt{1+4\lambda}}\right)+\lambda
z^2\ln{\left(\frac{2}{1+\sqrt{1+4\lambda}}\right)\ ,}
\end {equation}
and that for $\lambda \geq 0$ the following inequalities are valid:
$$1-\frac{2}{1+\sqrt{1+4\lambda}}\ \leq \ 1\ ,\
\frac{2}{1+\sqrt{1+4\lambda}}\ \leq \ \frac{1}{\sqrt{\lambda}}\ ,$$
we find the upper bound to remainder (\ref{rem3}) :
\begin {eqnarray}
&&\frac{M_n}{z^{2n}}\sum_{k_i=N_0+1}^{\infty}\left(\frac{\xi}{2}\right)^{2k_{i}}\left(k_i\right)^{k_{i+1}+k_{i-1}-1/2}
\exp{\left(\frac{k_{i+1}(k_{i+1}+1/2)+k_{i-1}(k_{i-1}+1/2)}{k_i}\right)}\exp{\left(k_i+\frac{k_{i+1}+k_{i-1}}{2}\right)}
\nonumber\\
&&\left(\frac{z}{k_i}\right)^{1/2}\frac{z^{2k_i+k_{i+1}+k_{i-1}}}{(k_i)^{k_i+\frac{k_{i+1}+k_{i-1}}{2}}}
\left(\frac{1}{\sqrt{1+\frac{k_{i+1}}{k_i}}}\right)^{k_{i+1}+k_i}
\left(\frac{1}{\sqrt{1+\frac{k_{i-1}}{k_i}}}\right)^{k_{i-1}+k_i}
\end {eqnarray}
For $k_i$ big, we use the identity:
\begin {equation}
\left(\frac{1}{\sqrt{1+\frac{k_{i\pm1}}{k_i}}}\right)^{k_{i\pm1}+k_i}\sim
\left(1-\frac{k_{i\pm1}}{k_i}\right)^{k_{i\pm1}+k_i}\sim
\exp{(-\frac{k_{i\pm1}}{2})}
\end {equation}
Finally, we have:
\begin {equation}
\frac{M_n}{z^{2n-k_{i+1}-k_{i-1}-1/2}}\sum_{k_i=N_0+1}^{\infty}\frac{1}{k_i!}\left(\frac{\xi
z}{2}\right)^{2k_{i}} \left(k_i\right)^{\frac{k_{i+1}+k_{i-1}-1}{2}}
\exp{\left(\frac{k_{i+1}(k_{i+1}+1/2)+k_{i-1}(k_{i-1}+1/2)}{k_i}\right)}
\end {equation}
The function

\begin {equation}
\exp{\left(\frac{k_{i+1}(k_{i+1}+1/2)+k_{i-1}(k_{i-1}+1/2)}{k_i}\right)}
\end {equation}
is decreasing for $k_i \geq N_0$, therefore following mean value
lemma we read:
\begin {equation}
\frac{M_n}{z^{2n-k_{i+1}-k_{i-1}-1/2}}\exp{(\rho_k)}\sum_{k_i=N_0+1}^{\infty}\frac{1}{k_i!}\left(\frac{\xi
z}{2}\right)^{2k_{i}} \left(k_i\right)^{\frac{k_{i+1}+k_{i-1}-1}{2}}
\end {equation}
where $\rho_k= (k_{i+1}(k_{i+1}+1/2)+k_{i-1}(k_{i-1}+1/2))/2,\
k\in(N_0,\infty).$ The series
\begin {equation}
\sum_{N_0+1}^{\infty}\frac{1}{k!}\left(\frac{\xi^2z^2}{4}\right)^k(k)^{(k_{i+1}+k_{i-1}-1)/2}
\end {equation}
is the remainder of the series uniformly convergent to the function
$$f(x)=(x\partial_x)^m\ e^x,$$
where $x=\frac{\xi^2z^2}{4},$ and $m\geq(k_{i+1}+k_{i-1}-1)/2.$ By
choosing the parameter $n$, which corresponds to number of the
decomposition terms taken into account of double asymptotic
expansion of the function $\mathcal{D}$, this contribution to the
remainder can be done as small as we need.

Now we are going to estimate the last part of the remainder,
corresponding to the difference of the series:
\begin {eqnarray}
&&\sum_{k_i=N_0+1}^{\infty}\
\frac{\left(\xi\right)^{2k_{i}}}{(2k_{i})!}
\Gamma(k_{i-1}+k_{i}+1/2) \Gamma(k_{i}+k_{i+1}+1/2)
\mathcal{D}_{-k_{i-1}-k_{i}-1/2}\;(z)\frac{\exp{(-\mathcal{A}z^2)}}{(1+4\lambda)^{1/4}}
\left[
\sum_{k=0}^{n-1}\frac{f_k(\lambda)}{z^{2k}}\right]-\nonumber\\
&-&\sum\limits_{j=0}^{\mathcal{J}}\; \frac{(-1)^j }{j!\;(2z^2)^j}\
\sum_{k_i=N_0+1}^{\infty}\
\frac{\left(\xi\right)^{2k_{i}}}{(2k_{i})!}
\Gamma(k_{i-1}+k_{i}+1/2) \Gamma(k_{i}+k_{i+1}+2j+1/2)
\mathcal{D}_{-k_{i-1}-k_{i}-1/2}\;(z) \label{rem33}
\end {eqnarray}
Following the identity (\ref{eg21}) we can recognize in the sums
over $k_i$ the remainders of the uniformly convergent series. These
remainders are small, if $\lambda = k_i^2 / z$ is the large
quantity. Due to the uniform convergence, we can change the order of
the summations and we read:
\begin {eqnarray}
&&\sum_{k_i=N_0+1}^{\infty}\
\frac{\left(\xi\right)^{2k_{i}}}{(2k_{i})!}
\Gamma(k_{i-1}+k_{i}+1/2) \Gamma(k_{i}+k_{i+1}+1/2)
\mathcal{D}_{-k_{i-1}-k_{i}-1/2}\;(z)\nonumber\\
&&\left\{\frac{\exp{(-\mathcal{A}z^2)}}{(1+4\lambda)^{1/4}} \left[
\sum_{k=0}^{n-1}\frac{f_k(\lambda)}{z^{2k}}\right]-
\sum\limits_{j=0}^{\mathcal{J}}\;
\frac{(-1)^j\pochh{k_{i}+k_{i+1}+1/2}{2j} }{j!\;(2z^2)^j}\right\}\
 \label{rem4}
\end {eqnarray}
The double asymptotic expansion reduces to Poincar\' e - type
expansion \cite {temme2} when $a$ is fixed after expanding the
quantities in expansion that depend on $\lambda = a/z^2$ for small
values of this parameter. In the difference
\begin {equation}
\left\{\frac{\exp{(-\mathcal{A}z^2)}}{(1+4\lambda)^{1/4}} \left[
\sum_{k=0}^{n-1}\frac{f_k(\lambda)}{z^{2k}}\right]-
\sum\limits_{j=0}^{\mathcal{J}}\;
\frac{(-1)^j\pochh{k_{i}+k_{i+1}+1/2}{2j} }{j!\;(2z^2)^j}\right\}
\end {equation}
all terms where $k, j\leq \min{(n, \mathcal{J})}$ cancel one another
and the rest terms are proportional or smaller then $z^{-2\min{(n,
\mathcal{J})}}$. As in the case of previous contributions to the
remainder, this means, that this part of the remainder can be done
as small as we need.

We shown, that we can evaluate $N-$ dimensional integral
(\ref{fin1}) by method of Poincar\' e decomposition of one of the
parabolic cylinder functions. We are able to accomplish the sum over
the summation index of the Taylor's decomposition of the linear
exponential function. We shown, that in the continuum limit the $N-$
dimensional integral evaluated by this method have the same value as
the corresponding continuum functional integral.

The result of the recurrence summation procedure of the principal
contribution is (see Appendix 1):

\begin {equation}
\mathcal{Z}_N\;=\;\left\{\prod\limits_{i=0}^{N}\;\left[2(1+b\triangle^2/c)\omega_
i\right]\right\}^{-1/2}
\;\sum\limits_{\mu=0}^{\mathcal{J}}\;\frac{(-1)^{\mu}}{\mu!\;(2z^2)^{\mu}}\;\left(N\right)_{2
\mu,\;0} \label{recu1}
\end {equation}

 Symbol
$\left(N\right)_{2j,\;i}$ is defined by the recurrence relation:
\begin {eqnarray}
\left(\alpha+1\right)_{2\mu,\;p} &=& \sum\limits_{\lambda=0}^{\mu}
\binom {\mu}{\lambda} \omega_{\alpha}^{-2(\mu-\lambda)}\;
\sum\limits^{2\lambda}_{i=max [ 0,\;p-2(\mu-\lambda)]}\;
\left(\frac{A^2}{\omega_{\alpha-1}\omega_{\alpha}}\right)^i\;
\left(\alpha\right)_{2\lambda,\;i}\;
a_p^{2(\mu-\lambda)+i}\nonumber\\
\noalign{\vskip8pt}
 \left(\Lambda=1\right)_{2\mu,\;p} &=& \frac{a_p^{2\mu}}{\omega_0^{2\mu}}
\end {eqnarray}
where
\begin {eqnarray}
 \omega_i\; &=& \;1-\frac{A^2}{\omega_{i-1}},\nonumber\\
 A &=&\frac{1}{2(1+b\triangle^2/c)},\nonumber\\
  \omega_0 &=& \frac {1}
{(1/2+b\triangle^2/c)}\ .
\end {eqnarray}

$\mathcal{Z}_N$ in the relation (\ref{recu1}) is the $N-th$
approximation of the functional integral. The continuum limit of
$\mathcal{Z}_N$ we calculate by the procedure proposed by Gelfand-
Yaglom for the calculation of the functional integral for the
harmonic oscillator \cite{bkz}. The method and the solutions are
discussed in the Appendix 3, here we present the result only.

The functional integral in the continuum limit is defined by the
relation
$$\lim_{N->\infty}\, \mathcal{Z}_N = \frac{1}{\sqrt{F(\beta)}}$$
where the function $F(\beta)$ is the solution of the differential
equation taken in the point $\beta$:
\begin {equation}
\frac{\partial^2}{\partial
\tau^2}F(\tau)+4\frac{\partial}{\partial \tau}F(\tau)\,
\frac{\partial}{\partial \tau}\ln{S(\tau)}=
F(\tau)\left(\frac{2b}{c}-2\frac{\partial^2}{\partial
\tau^2}\ln{S(\tau)-4\left(\frac{\partial}{\partial \tau}\ln{S(\tau)}\right)^2}\right) \label{gyeq1}
\end {equation}
The initial conditions are:
\begin {eqnarray}
F(0) &=& 1,
\label{gyeq2} \\
\frac{\partial}{\partial \tau}F(0) &=& 0.\nonumber
\end{eqnarray}
The function $S(\tau)$ is defined as:
\begin {equation}
S(\tau)=\lim_{\triangle \rightarrow 0}\,
\sum\limits_{\mu=0}^{\mathcal{J}}\;\frac{(-1)^{\mu}}{\mu!\;(2z^2\triangle^{3})^{\mu}}\;
\left(\triangle^{3\mu}(N)_{2 \mu,\;0}\right)
\end {equation}

The equation (\ref{gyeq1}) can be simplified by substitution
$$F(\tau) = \frac{y(\tau)}{(S(\tau))^2}$$
For $y(\tau)$ we find the equation:
\begin {equation}
\frac{\partial^2}{\partial \tau^2}y(\tau) =
y(\tau)\ \frac{2b}{c}
\end {equation}

In the present time the calculation of the function $S(\tau)$ is
not finished, therefore we present the calculation of the lowest
order term in the power expansion in the coupling constant in the
Appendix 2. For the first two terms of such expansion in the
continuum we find:
$$\lim_{\triangle \rightarrow 0}\sum\limits_{\mu=0}^{1}\;
\frac{(-1)^{\mu}}{\mu!\;(2z^2)^{\mu}}\;\left(N\right)_{2
\mu,\;0} = 1 -\frac{\pochh{1/2}{2}}{2}\; \frac{a}{c^2\gamma^3}\;
\left\{\tanh(\tau\gamma) + \tau\gamma\left[
3\tanh^2(\tau\gamma)-1\right]\right\}$$ where $$\gamma
=\sqrt{\frac{2b}{c}}\; .$$

\section*{Conclusions}

In the present article we presented the new method of the
calculation of the Wiener unconditional measure functional
integral for the action with the fourth order term. This method
can be extended to the case with $\varphi^{2n}$, $n=3, 4, ...$
term in the action, but we did not discuss these possibilities
here. The main results are the analytic formula for the $N$
dimensional integral and the generalized Gelfand-Yaglom equation
implying the functional integral in the continuum limit.

\vskip 0.3cm {\bf{Acknowledgements}}. This work was supported by
VEGA projects No. 2/3106/2003 (J.B.) and No. 1/025/03 (P.P). The
authors acknowledge J. Polonyi and M. Znojil for the discussions
and interest about this work.

\section*{Appendix 1}
\appendix
In this appendix we propose and on pedagogical level to evaluate the $k_i$ summations by the
reccurence method.
 Let we start with summation over the index
$k_{N-1}$ of the Eq.(\ref{fin1}). The sum that must be done is:
\begin {eqnarray}
\mathcal{Z}_1 &=& \sum \limits_{k_{N-1}=0}^\infty \;
\left[\frac{1}{\sqrt{2\pi(1+b\triangle^2/c)}} \;
\frac{\left(\frac{\scriptstyle \xi^2}{\scriptstyle \omega_0}\right)^{k_{N-1}}}{(2k_{N-1})!}
\Gamma(k_{N-2}+k_{N-1}+1/2)\mathcal{D}_{-k_{N-2}-k_{N-1}-1/2}\;(z)\right]\
 \nonumber \\
& & \left[\frac{1}{\sqrt{2\pi(1+b\triangle^2/c)\omega_0}} \;
\Gamma(k_{N-1}+1/2)\mathcal{D}_{-k_{N-1}-1/2}\;(z_0)\right]
\label{app2}
\end {eqnarray}
Let us remind the dependence of the function $\mathcal{D}$ on the
parabolic cylinder function $D$:
$$\mathcal{D}_{-m-1/2}(z) = z^{m+1/2}\;
e^{\frac{\scriptstyle z^2}{\scriptstyle 4}}\; D_{-m-1/2}(z)$$ and,
also the definitions of the variables $\xi$, $\omega_0$, $z$ and $z_0$:
\begin {eqnarray*}
z & = & \frac{c(1+b\triangle^2/c)}{\sqrt{2a\triangle^3}}\\
z_0 & = & \frac{c(1/2+b\triangle^2/c)}{\sqrt{2a\triangle^3}}\\
\xi & =&  \frac{1}{1+b\triangle^2/c}\\
\omega_0 &=& \frac{1/2+b\triangle^2/c}{1+b\triangle^2/c}\ .
\end {eqnarray*}
The sum (\ref{app2}) is uniformly convergent, therefore for an
arbitrary small positive number $\varepsilon$ exist the number $N_0$
such that the following inequality is true:
\begin {equation}
\mid \sum\limits_{k_{N-1}=N_0+1}^\infty \;\mid \; < \;
N^{-1-\varepsilon}
\end {equation}
When we replace the true sum by the truncated one, we introduce into
our calculation error up to the order $N^{-1-\varepsilon}$. All
recurrence procedure consists of the $N$ steps, therefore we require
that in the continuum limit the influence of the reminders of the
sums disappears.  For the truncated sum we demand that
 $$N_0\; <\; z\; \approx \; N^{3/2}$$
Therefore we are permitted to use the asymptotic Poincar\' e type
expansion of the parabolic cylinder function, which for
$\mathcal{D}$ means:
\begin {equation}
\mathcal{D}_{-k_n-1/2}(z_0)\; \equiv
\; z_0^{k_n+1/2}\;e^{z_0^2/4}\;D_{-k_n-1/2}(z_0)\;=
\;\sum\limits_{j=0}^{\mathcal{J}}\; (-1)^j \;\frac{\pochh{k_n+1/2}{2j}}{j!\;(2z_0^2)^j}
\end {equation}
In the last relation, $\mathcal{J}$ means the number of the terms of
the asymptotic expansions convenient to take into account. We apply
asymptotic expansion for the function
$\mathcal{D}_{-k_{N-1}-1/2}\;(z_0)$ in Eq.(\ref{app2}). In truncated
sum we interchange the order of the finite summations over indices
$k_{N-1}$ and $j$. We replace $\mathcal{D} $ by $D$, therefore a
corresponding power of the variable $z$ play important role.   We
obtained the relation:
\begin {eqnarray}
\mathcal{Z}_1^{cut}
&=&\frac{\Gamma{(1/2)}}{\sqrt{2\pi(1+b\triangle^2/c)\omega_0}}\;\sum\limits_{j=0}^{\mathcal{J}}\;
\frac{(-1)^j}{j!\;(2z_0^2)^j}\;
\frac{\exp{(z^2/4)}}{\sqrt{2\pi(1+b\triangle/c)}} z^{k_{N-2}+1/2}
\\&& \sum\limits_{k_{N-1}=0}^{N_0}\frac{(\frac{\scriptstyle z\
\xi^2}{\scriptstyle 4\omega_0})^{k_{N-1}}}{(k_{N-1})!}\;
\Gamma(k_{N-1}+k_{N-2}+1/2)\;\pochh{k_{N-1}+1/2}{2j}\;
D_{-k_{N-1}-k_{N-2}-1/2}\;(z)\nonumber \label{z1}
\end {eqnarray}
Here we simplified the calculations by identity:
$$(2k)! = 2^{2k}\; k!\; \pochh{1/2}{k},\ \Gamma{(k+1/2)}=\Gamma{(1/2)}\pochh{1/2}{k}$$

Let us study in details the sum
\begin {equation}
\sum\limits_{k_{N-1}=0}^{N_0}\frac{(\frac{z\
\xi^2}{4\omega_0})^{k_{N-1}}}{(k_{N-1})!}\;
\Gamma(k_{N-1}+k_{N-2}+1/2)\;\pochh{k_{N-1}+1/2}{2j}\;
D_{-k_{N-1}-k_{N-2}-1/2}\;(z)
\end {equation}
This sum is uniformly convergent, as we could prove by the same
procedure as in section "Summation over $k_i$". Therefore we can
extend the summation up to infinity with the error of the order
$N^{-1-\varepsilon}$. To be able provide the sum over index
$k_{N-1}$, we must modify term $$\pochh{k_{N-1}+1/2}{2j}\ .$$ We
see, that this object is a polynomial in variable $k_{N-1}$ of the
$2j-th$ order. We rewrite the polynomial in the another form:
$$\pochh{k_{N-1}+1/2}{2j}\; = \;
\sum\limits_{i=0}^{\min{(2j,k_{N-1})}}a_i^{2j}\frac{(k_{N-1})!}{(k_{N-1}-i)!}$$
The coefficients $a_i^{2j}$ can be calculated by
recurrence procedure from the relation:
\begin {equation}
\sum\limits_{k_{N-1}=0}^{N_0}\frac{(k_{N-1}+1/2)_{2j}}{(k_{N-1})!}\;f(k_{N-1})\;
=\sum\limits_{i=0}^{2j}\;a_i^{2j}\;
\sum\limits_{k_{N-1}=i}^{N_0}\frac{1}{(k_{N-1}-i)!}\;f(k_{N-1})
\end {equation}
From the above definition, we find the recurrence equation:
\begin {equation}
a^k_i=(k-1/2+i)a^{k-1}_i\; + a^{k-1}_{i-1}
\end {equation}
when the initial conditions are: $$a_j^j=1\;,\ a^j_0 =
\pochh{1/2}{j}\; ,\ a^j_{j+1}=0$$ The solution of this recurrence
equation is:
\begin {equation}
a_i^{j}\; = \;\binom{j}{i}\frac{\pochh{1/2}{j}}{\pochh{1/2}{i}}
\end {equation}
Inserting all these replacements into Eq.(\ref{z1}), with help of
the identity
$$\Gamma(k_{N-2}+k_{N-1}+1/2) = \Gamma(k_{N-2}+i+1/2)\;
\pochh{k_{N-2}+i+1/2}{k_{N-1}-i}$$ after some algebra, by
introducing the new summation index $k=k_{N-1} - i $ we obtain the
formula:
\begin {eqnarray}
\mathcal{Z}_1^{cut}
&=&\frac{z^{k_{N-2}+1/2}}{\sqrt{2(1+b\triangle^2/c)\omega_0}}\;\sum\limits_{j=0}^{\mathcal{J}}\;
\frac{(-1)^j}{j!\;(2z^2_0)^j}\;
\frac{e^{z^2/4}}{\sqrt{2\pi(1+b\triangle^2/c)}}
\sum\limits_{i=0}^{2j}\;a_i^{2j}\\
&\times&
\Gamma(k_{N-2}+i+1/2)\;\left(\frac{z\ \xi^2}{4\omega_0}\right)^i\;\sum\limits_{k=0}^{N_0}
\frac{(\frac{z\ \xi^2}{4\omega_0})^k}{k!}\;(k_{N-2}+i+1/2)_{k}\;
D_{-k_{N-2}-i-1/2-k}\;(z)\nonumber \quad
\end {eqnarray}
In the above relation we extend the summation over the index $k$ up
to infinity, because the added reminder disappear in the continuum
limit. The sum over $k$ is now prepared for application of the
identity:
\begin {equation}
e^{x^2/4}\sum\limits_{k=0}^{\infty}\;
\frac{\pochh{\nu}k}{k!}\;t^k\;D_{-\nu-k}(x)\;=\;e^{(x-t)^2/4}\;D_{-\nu}\;(x-t)
\end {equation}
The result of the first  recurrence step, replacing $D$ by
$\mathcal{D}$, can be read:
\begin {equation}
\mathcal{Z}_1^{cut} =\frac{1}{\sqrt{2(1+b\triangle^2/c)\omega_0}}
\frac{(\omega_1)^{-k_{N-2}}}{\sqrt{2\pi(1+b\triangle^2/c)\omega_1}}
\sum\limits_{j=0}^{\mathcal{J}}\frac{(-1)^j}{j!\;(2z_0^2)^j}
\sum\limits_{i=0}^{2j}a_i^{2j}\left(\frac{\xi^2}{4\omega_0 \omega_1}\right)^i
\Gamma(k_{N-2}+i+1/2)\mathcal{D}_{-k_{N-2}-i-1/2}\;(z_1)
\label{rez1}
\end {equation}
where $$z_1=z\left(1-\xi^2/(4 \omega_0)\right),\; \omega_1 = \frac{z_1}{z} = 1-\xi^2/(4 \omega_0)\; .$$

For the second recurrence step, we have the summation over the
index $k_{N-2}$ in the form:
\begin {equation}
\mathcal{Z}_2=\sum \limits_{k_{N-2}=0}^\infty \;
\left[\frac{1}{\sqrt{2\pi(1+b\triangle^2/c)}} \;
\frac{(\xi)^{2k_{N-2}}}{(2k_{N-2})!}
\Gamma(k_{N-3}+k_{N-2}+1/2)\mathcal{D}_{-k_{N-3}-k_{N-2}-1/2}\;(z)\right]\
\mathcal{Z}_1^{cut} \label{app3}
\end {equation}
Before replacing $\mathcal{Z}_1^{cut}$ in the last equation, let
us define the first step of the new recurrence relation by:
$$\left(1\right)_i^{2j} = \frac{1}{\omega_0^{2j}}a_i^{2j}$$
Then the above equation, taking into account the identity $z_0 = z \omega_0$ in detail can be read:
\begin {eqnarray}
\mathcal{Z}_2 &=& \frac{1}{\sqrt{2(1+b\triangle^2/c)\omega_0}}
\frac{1}{\sqrt{2(1+b\triangle^2/c)\omega_1}} \\
& \times & \sum\limits_{k_{N-2}=0}^\infty \;
\left[\frac{1}{\sqrt{2\pi(1+b\triangle^2/c)}} \;
\frac{\left(\frac{\scriptstyle \xi^2}{\scriptstyle
\omega_1}\right)^{k_{N-2}}}{(2k_{N-2})!}
\Gamma(k_{N-3}+k_{N-2}+1/2)\mathcal{D}_{-k_{N-3}-k_{N-2}-1/2}\;(z)\right]\nonumber\\
& \times &
\sum\limits_{j=0}^{\mathcal{J}}\;\frac{(-1)^j}{j!\;(2z^2)^j}\;
\sum\limits_{i=0}^{2j}\;\left(1\right)_i^{2j}\;
\left(\frac{\xi^2}{4 \omega_0 \omega_1}\right)^i\;
\Gamma(k_{N-2}+i+1/2)\mathcal{D}_{-k_{N-2}-i-1/2}\;(z_1)\nonumber
\label{app4}
\end {eqnarray}
Now, due to the uniform convergence of the sum over index $k_{N-2}$
we will work with truncated sum of the complete $\mathcal{Z}_2$. In
the truncated sum, we use the asymptotic Poincar\' e type expansion
of the parabolic cylinder function
$\mathcal{D}_{-k_{N-2}-i-1/2}\;(z_1)$. We have then in the relation
for $\mathcal{Z}_2^{cut}$ the finite summations only, so we
rearrange them, in order that summation over $k_{N-2}$ will be
provided first. Remember, that in an intermediate step we replace
$\mathcal{D}_{-k_{N-3}-k_{N-2}-1/2}\;(z)$ by
$D_{-k_{N-3}-k_{N-2}-1/2}\;(z)$ and we must take into account the
corresponding power of variable $z$. We have:
\begin {eqnarray}
\mathcal{Z}_2^{cut}&=&
\frac{1}{\sqrt{2(1+b\triangle^2/c)\omega_0}}\;
\frac{1}{\sqrt{2(1+b\triangle^2/c)\omega_1}}
\frac{1}{\sqrt{2\pi(1+b\triangle^2/c)}} \\
& \times &
\sum\limits_{j=0}^{\mathcal{J}}\;\frac{(-1)^j}{j!\;(2z^2)^j}\;
\sum\limits_{i=0}^{2j}\;\left(1\right)_i^{2j}\;
\left(\frac{A^2}{\omega_0\omega_1}\right)^i\;
\sum\limits_{l=0}^{\mathcal{L}}\;\frac{(-1)^l}{l!\;(2z^2)^l}
\frac{1}{\omega_1^{2l}}\; e^{z^2/4}\ z^{k_{N-3}+1/2}\nonumber\\
 & \times & \sum\limits_{k_{N-2}=0}^{N_0} \;
\frac{\left(\frac{z\ \xi^2}{4 \omega_1}\right)^{k_{N-2}}}{k_{N-2}!}
\Gamma(k_{N-3}+k_{N-2}+1/2)\ D_{-k_{N-3}-k_{N-2}-1/2}\;(z)\
\pochh{k_{N-2+1/2}}{2l+i}\; ,\nonumber
\end {eqnarray}
if the identity $$A = \frac{\xi}{2}$$ was introduced. Summing over index $k_{N-2}$ as in the first recurrence step, we
have:
\begin {eqnarray}
\mathcal{Z}_2^{cut}&=&
\frac{1}{\sqrt{2(1+b\triangle^2/c)\omega_0}}\;
\frac{1}{\sqrt{2(1+b\triangle^2/c)\omega_1}}
\frac{(\omega_2)^{-k_{N-3}}}{\sqrt{2(1+b\triangle^2/c)\omega_2}}
\sum\limits_{j=0}^{\mathcal{J}}\;\sum\limits_{l=0}^{\mathcal{L}}\;
\frac{(-1)^{j+l}}{j!l!(2z^2)^{j+l}}\;\frac{1}{\omega_1^{2l}}\\
& \times & \sum\limits_{i=0}^{2j}\;\left(1\right)_i^{2j}\;
\left(\frac{A^2}{\omega_0\omega_1}\right)^i\;
\sum\limits_{p=0}^{2l+i}\;a_p^{2l+i}\left(\frac{A^2}{\omega_1\omega_2}\right)^p\;
\Gamma(k_{N-3}+p+1/2)\mathcal{D}_{-k_{N-3}-p-1/2}(z_2)\; ,\nonumber
\end {eqnarray}
where one postulate the identities: $$z_2 = z \omega_2,\; \omega_2 = 1-\frac{A^2}{\omega_1}\; .$$

By adjustment in the summations:
$$\sum\limits_{j=0}^{\mathcal{J}}\;\sum\limits_{l=0}^{\mathcal{L}}\;
\frac{(-1)^{j+l}}{j!l!(2z^2)^{j+l}}\;(\cdots j,\;l \cdots) =
\;\sum\limits_{\mu=0}^{\mathcal{J}+\mathcal{L}}\frac{(-1)^{\mu}}{\mu!(2z^2)^{\mu}}\;
\sum\limits_{j=0}^{\mu}\binom{\mu}{j}(\cdots j,\;l=\mu-j \cdots)$$
and interchange of the order of the summations:
$$\sum\limits_{i=0}^{2j}\;\sum\limits_{p=0}^{2\mu-2j+i}\;
\rightarrow \sum\limits_{p=0}^{\mu}\;
\sum\limits^{2j}_{i=\max{[0,\; p-2\mu+2j]}}$$ we find the result
of the second recurrence step:
\begin {eqnarray}
\mathcal{Z}_2^{cut}&=&
\frac{1}{\sqrt{2(1+b\triangle^2/c)\omega_0}}\;
\frac{1}{\sqrt{2(1+b\triangle^2/c)\omega_1}}
\frac{(z/z_2)^{k_{N-3}}}{\sqrt{2(1+b\triangle^2/c)\omega_2}}\\
&\times&
\sum\limits_{\mu=0}^{\mathcal{J}+\mathcal{L}}\frac{(-1)^{\mu}}{\mu!(2z^2)^{\mu}}\;
\sum\limits_{p=0}^{2\mu}\;\left(2\right)_p^{2\mu}\left(\frac{A^2}{\omega_1\omega_2}\right)^p
\Gamma(k_{N-3}+p+1/2)\mathcal{D}_{-k_{N-3}-p-1/2}(z_2)\nonumber
\end {eqnarray}
Where the second recurrence step of the function is defined:
\begin {equation}
\left(2\right)_p^{2\mu}=
\sum\limits_{j=0}^{\mu}\;\binom{\mu}{j}\frac{1}{\omega_1^{2\mu-2j}}
\sum\limits^{2j}_{i=\max{[0,\; p-2\mu+2j]}}
\left(\frac{A^2}{\omega_0\omega_1}\right)^i\;
\left(1\right)_i^{2j}\; a_p^{2\mu-2j+i}
\end {equation}
We can see, that after the $\Lambda$ recurrence steps, the result
of the $\Lambda$ summations over the indices $k_i$ can be read:
\begin {eqnarray}
\mathcal{Z}_{\Lambda}^{cut}&=&
\left\{\prod\limits_{i=0}^{\Lambda}\frac{1}{\sqrt{2(1+b\triangle^2/c)\omega_i}}\right\}
(\omega_{\Lambda})^{-k_{N-1-\Lambda}} \label {app11}\\ &\times&
\sum\limits_{\mu=0}^{\mathcal{J}_1+\cdots
+\mathcal{J}_\Lambda}\frac{(-1)^{\mu}}{\mu!(2z^2)^{\mu}}\;
\sum\limits_{p=0}^{2\mu}\;\left(\Lambda\right)_p^{2\mu}
\left(\frac{A^2}{\omega_{\Lambda-1}\omega_{\Lambda}}\right)^p
\Gamma(k_{N-\Lambda-1}+p+1/2)\mathcal{D}_{-k_{N-\Lambda-1}-p-1/2}(z_{\Lambda})\; .\nonumber
\end {eqnarray}
We have postulated the recurrence relations: $$z_{\Lambda} = z \omega_{\Lambda}, \;
\omega_{\Lambda} = 1-\frac{A^2}{\omega_{\Lambda-1}},\; \omega_0 = \frac{1/2 + b\triangle^2/c}{1 + b\triangle^2/c}\; ,$$
and evaluated the recurrence definition for the function
$\left(\Lambda\right)_p^{2\mu}$:
\begin {equation}
\left(\Lambda\right)_p^{2\mu}=
\sum\limits_{j=0}^{\mu}\;\binom{\mu}{j}\frac{1}{\omega_{\Lambda-1}^{2\mu-2j}}
\sum\limits^{2j}_{i=\max{[0,\; p-2\mu+2j]}}
\left(\frac{A^2}{\omega_{\Lambda-2}\omega_{\Lambda-1}}\right)^i\;
\left(\Lambda-1\right)_i^{2j}\; a_p^{2\mu-2j+i}
\end {equation}
when the recurrence procedure begin from:
$$\left(1\right)_i^{2j} = \frac{1}{\omega_0^{2j}}a_i^{2j}$$

 After the last
recurrence step, for $\Lambda=N-1$, we are left with the relation
of the form (\ref {app11}) where $k_{N-\Lambda-1}\equiv 0$ and the
index of the $\mathcal{D}$ function is only $-p-1/2$. We expand
the $D_{-p-1/2}(z_{N-1})$ as in all previous recurrence steps and
we find the relation:
\begin {eqnarray}
\mathcal{Z}_{N-1}^{cut}&=&
\left\{\prod\limits_{i=0}^{N-1}\frac{1}{\sqrt{2(1+b\triangle^2/c)\omega_i}}\right\}
\sum\limits_{\mu=0}^{\mathcal{J}_1+\cdots
+\mathcal{J}_N}\frac{(-1)^{\mu}}{\mu!(2z^2)^{\mu}}\nonumber \\
 &\times&
\sum\limits_{l=0}^{\mu}\;\binom{\mu}{l}\frac{1}{\omega_{N-1}^{2l}}
\sum\limits^{2l}_{i=0}
\left(\frac{A^2}{\omega_{N-2}\omega_{N-1}}\right)^i\;
\left(N-1\right)_i^{2j}\;\pochh{1/2}{2l+i}
\end {eqnarray}
Following the definition of the $a_i^j$ symbols, we have:
$$\pochh{1/2}{2l+i} =  a_0^{2l+i}$$ Then in the last part of the
preceding equation we read:
\begin {equation}
\sum\limits_{l=0}^{\mu}\;\binom{\mu}{l}\frac{1}{\omega_{N-1}^{2l}}
\sum\limits^{2l}_{i=0}
\left(\frac{A^2}{\omega_{N-2}\omega_{N-1}}\right)^i\;
\left(N-1\right)_i^{2j}\;a_0^{2l+i}\; = \; \left(N\right)_0^{2\mu}
\end {equation}

Following the calculations done in this Appendix, we conclude,
that it is possible to provide the summations in the precise
formula for the $N$ dimensional integral at least by help of the
asymptotic expansions of the parabolic cylinder functions. It is
possible to provide the continuum limit of the our result and
there are no additional terms contributing to the result in the
continuum limit. The result can be read
\begin {equation}
\mathcal{Z}_{N-1}^{cut}=
\left\{\prod\limits_{i=0}^{N-1}\frac{1}{\sqrt{2(1+b\triangle^2/c)\omega_i}}\right\}
\sum\limits_{\mu=0}^{\mathcal{J}_1+\cdots
+\mathcal{J}_N}\frac{(-1)^{\mu}}{\mu!(2z^2)^{\mu}}\left(N\right)_0^{2\mu}
\label{vysl1}
\end {equation}
This expression is sufficient for  calculation of the continuum
Wiener unconditional measure functional integral by Ge\`lfand-Yaglom
procedure leading to the differential equation of the second order.
The second part of relation (\ref{vysl1}) represent the expansion of
an unknown function in present time.

\section*{Appendix 2}
\appendix
In this appendix we will present the idea of the decomposition of
the principal result by the summations over the indices $k_i$ by
slightly different method as was presented in the Appendix 1. Our
goal is to study the unknown function from Appendix 1 in the power
expansion in the coupling constant. We start as in Appendix 1 from
the precise relations for the $N$ dimensional functional integral,
and we proceed by the recurrence procedure withouth the
introducing the function $(\Lambda)_i^{2j}$. We start, as in
Appendix 1, by summation over the index $k_{N-1}$:
\begin {eqnarray}
\mathcal{Z}_1 &=& \sum \limits_{k_{N-1}=0}^\infty \;
\left[\frac{1}{\sqrt{2\pi(1+b\triangle/c)}} \;
\frac{\left(\frac{\scriptstyle \xi^2}{\scriptstyle z\;
z_0}\right)^{k_{N-1}}}{(2k_{N-1})!}
\Gamma(k_{N-2}+k_{N-1}+1/2)\mathcal{D}_{-k_{N-2}-k_{N-1}-1/2}\;(z)\right]\nonumber
\\ && \left[\frac{1}{\sqrt{\frac{2\pi}{c}\; z_0\; \sqrt{2a\triangle^3}}}
\; \Gamma(k_{N-1}+1/2)\mathcal{D}_{-k_{N-1}-1/2}\;(z_0)\right]
\label{aps2}
\end {eqnarray}
Repeating the same calculations as before, we will have after the
second recurrence step the relations:
\begin {eqnarray}
\mathcal{Z}_2^{cut}&=&
\frac{1}{\sqrt{2(1+b\triangle^2/c)\omega_0}}\;
\frac{1}{\sqrt{2(1+b\triangle^2/c)\omega_1}}
\frac{(z/z_2)^{k_{N-3}+1/2}}{\sqrt{2(1+b\triangle^2/c)}}
\sum\limits_{j_1=0}^{\mathcal{J}_1}\;\sum\limits_{j_2=0}^{\mathcal{J}_2}\;
\frac{(-1)^{j_1+j_2}}{j_1!j_2!(2z^2)^{j_1+j_2}}\;\frac{1}{\omega_0^{2j_1}\omega_1^{2j_2}}\\
& \times & \sum\limits_{i_1=0}^{2j_1}\;a_{i_1}^{2j_1}\;
\left(\frac{\xi^2}{4z_0z_1}\right)^{i_1}\;
\sum\limits_{i_2=0}^{2j_2+i_1}\;a_{i_2}^{2j_2+i_1}\left(\frac{\xi^2}{4z_1z_2}\right)^{i_2}\;
\Gamma(k_{N-3}+i_2+1/2)\mathcal{D}_{-k_{N-3}-i_2-1/2}(z_2)\nonumber
\end {eqnarray}
In contrary to the method in the Appendix 1, we don't combine the
summations over the asymptotic expansion indices. We see, that by
this procedure we replace the summations over indices $k_i$ by
summations over the asymptotic expansions indices $j_i$. Taking
thoroughly into account the last summation term, we find:
\begin {eqnarray}
\mathcal{Z}_N^{cut} &=&
\left\{\prod\limits_{i=1}^N[2(1+b\triangle^2/c)\omega_i]\right\}^{-1/2}\
\sum\limits_{j_1,\cdots,j_N}^{\mathcal{J}_1,\cdots,\mathcal{J}_N}\left[\prod\limits_{i=1}^N\;
\frac{(-1)^{j_i}}{(j_i)!(2z^2)^{j_i}\omega_{i-1}^{2j_i}}\right]\nonumber\\
&\times &
\sum\limits_{i_1=0}^{2j_1}\;\binom{2j_1}{i_1}\frac{\pochh{1/2}{2j_1}}{\pochh{1/
2}{2i_1}}
\left(\frac{A^2}{\omega_0 \omega_1}\right)^{i_1}\nonumber\\
&\times & \cdots \nonumber\\
&\times &
\sum\limits_{i_\mu=0}^{2j_\mu+i{\mu-1}}\;\binom{2j_{\mu}+i_{\mu-1}}{i_\mu}
\frac{\pochh{1/2}{2j_\mu+i{\mu-1}}}{\pochh{1/2}{2i_\mu}}
\left(\frac{A^2}{\omega_{\mu-1} \omega_\mu}\right)^{i_\mu}\\
&\times & \cdots \nonumber\\
&\times &
\sum\limits_{i_{N-1}=0}^{2j_{N-1}+i_{N-2}}\;\binom{2j_{N-1}+i_{N-2}}{i_{N-1}}
\frac{\pochh{1/2}{2j_{N-1}+i_{N-2}}}{\pochh{1/2}{2i_{N-1}}}
\left(\frac{A^2}{\omega_{N-2} \omega_{N-1}}\right)^{i_{N-1}}
\times \pochh{1/2}{2i_{N-1}}\pochh{1/2+i_{N-1}}{2j_{N}}\nonumber\\
\end {eqnarray}
In the last equation the identities were used:
$$\frac{\xi^2}{4z_{i-1}z_i}\; =\; \frac{A^2}{\omega_{i-1}
\omega_i}\ ,$$ where $$A=\frac{\xi}{2z}=\frac{1}{2(1+b
\triangle^2/c)}$$ and
$$a^k_i\; =\; \binom{k}{i}\frac{\pochh{1/2}{k}}{\pochh{1/2}{i}}$$
By the useful identity $$ \pochh{1/2}{2j_{\mu}+i_{\mu-1}}=
\pochh{1/2}{i_{\mu-1}}\;\pochh{1/2+i_{\mu-1}}{2j_{\mu}}$$ we
convert our result to the more reliable form for the consecutive
calculations:
\begin {eqnarray}
\mathcal{Z}_N^{cut} &=&
\left\{\prod\limits_{i=1}^N[2(1+b\triangle^2/c)\omega_i]\right\}^{-1/2}\
\sum\limits_{j_1,\cdots,j_N}^{\mathcal{J}_1,\cdots,\mathcal{J}_N}\left[\prod\limits_{i=1}^N\;
\frac{(-1)^{j_i}}{(j_i)!(2z^2)^{j_i}\omega_{i-1}^{2j_i}}\right]\;
\pochh{1/2}{2j_1}\nonumber\\
&\times &
\sum\limits_{i_1=0}^{2j_1}\;\binom{2j_1}{i_1}\pochh{1/2+i_1}{2j_2}
\left(\frac{A^2}{\omega_0 \omega_1}\right)^{i_1}\nonumber\\
&\times & \cdots \nonumber\\
&\times &
\sum\limits_{i_\mu=0}^{2j_\mu+i_{\mu-1}}\;\binom{2j_\mu+i_{\mu-1}}{i_\mu}
\pochh{1/2+i_ {\mu}}{2j_{\mu+1}} \left(\frac{A^2}{\omega_{\mu-1}
\omega_\mu}\right)^{i_\mu}
\label{aps3}\\
&\times & \cdots \nonumber\\
&\times &
\sum\limits_{i_{N-1}=0}^{2j_{N-1}+i_{N-2}}\;\binom{2j_{N-1}+i_{N-2}}{i_{N-1}}
\pochh{1/2+i_{N-1}}{2j_N} \left(\frac{A^2}{\omega_{N-2}
\omega_{N-1}}\right)^{i_{N-1}} \nonumber
\end {eqnarray}

In the calculation as follows the key role plays the objects
$\omega_i$ defined by recurrence relation $$\omega_i\; =\;
1-\frac{A^2}{\omega_{i-1}}$$ with the first term
$$\omega_0 \; =\; 1/2 + B$$ where
$$B\; = \; \frac{b\triangle^2/c}{2(1+b\triangle^2/c)}$$
The $\omega_i$ such defined are represented by the continued
fraction. The continued fraction can be represented by the simpler
relation as the solution of the $n-th$ convergent problem of the
continued fraction \cite{findif}. Let us shortly explain this
procedure.

Let us have a continued fraction of the form:
$$ \omega_n = a_1+\frac{\scriptstyle b_1}{\scriptstyle a_2+
\frac{\scriptstyle b_2}{\scriptstyle a_3+\cdots}}$$ The $n-th$
convergent is defined as
$$\omega_n = \frac{p_n}{q_n}$$
where $p_n$ and $q_n$ are defined by equations:
\begin {eqnarray}
p_n &=& a_n \;p_{n-1}+b_n \; p_{n-2}\nonumber\\
q_n &=& a_n \;q_{n-1}+b_n \; q_{n-2}\nonumber\\
\end {eqnarray}
The solutions of these recurrence equations have the form:
\begin {eqnarray}
p_n &=& \tilde{w}_1 \rho_1^n + \tilde{w}_2 \rho_2^n\nonumber\\
q_n &=& w_1 \rho_1^n + w_2 \rho_2^n \nonumber\\
\end {eqnarray}
where $\rho_{1,2}$ are the solutions of the characteristic
equation, in our case homogenous one:
$$ \rho^2 -a_n\rho-b_n = 0 $$
For the continued fraction  we have:
$$a_n=1$$
 $$b_n=-A^2$$ and the solution of the characteristic equation
 would be:
 $$\rho_{1,2} = \frac{1}{2}(1\pm \sqrt{1-4A^2}\;)$$
 The constants $\tilde{w}_1$ and $w_1$ are fixed by $\omega_0$ and
 $\omega_1$ terms, that adjust the initial conditions:
\begin {eqnarray*}
p_0 &=& 1+2B\\ p_1 &=& 1+2B-A^2\\ q_0 &=& 2\\ q_1 &=& 1+2B
\end {eqnarray*}
The $n-th$ convergent method solution is completed by the
relations:
\begin {eqnarray*}
\tilde{w}_{1,2} &=& \frac{1}{2}\left[(1+2B)\pm
\left(\sqrt{1-4A^2}+\frac{2B}{\sqrt{1-4A^2}}\right)\right]\\
w_{1,2} &=& 1 \pm \frac{2B}{\sqrt{1-4A^2}}
\end {eqnarray*}
The very important characteristic follows from the above solution:
$$p_n\; = \; q_{n+1}$$
which simplify our calculation significantly.

We are seeking for an expansion of the functional determinant. It
look reasonable to arrange the relation (\ref{aps3}) following the
condition $$\sum\; j_i\; =K =\; constant$$ where by the value of
this $constant$ is labeled the term composed by the sum of all the
terms of (\ref{aps3}) suitable satisfying to the above condition.
We can prove immediately, that if this $constant$, let us choose
$K$, is $K=0$ the contributing term is only one, it is the
relation where all the $j_i=0$ and value of this contribution is
$1$. Let us study the nontrivial case, when $K=1$. The
contributions will be all the members of the multiple sum
(\ref{aps3}) where only one index $j_i$ is nonzero and equal to
one. By some algebra, we can express all the contributions as the
sum:
\begin {equation}
\mathcal{Z}_N^1
=\left\{\prod\limits_{i=1}^N[2(1+b\triangle^2/c)\omega_i]\right\}^{-1/2}\
\mathcal{S}_N
\label{vysl3}
\end {equation}
where
\begin {eqnarray}
&\mathcal{S}_N&(j_1+j_2+\cdots+j_N=1)= 1 + \frac{-1}{1!\;2z^2}\pochh{1/2}{2} \label{aps4}\\
& \times & \sum\limits_{k=1}^{N-1}\frac{1}{\omega_{k-1}^2}\;
\sum\limits_{m=k}^{N-1}\;\sum\limits_{l=m}^{N-1}(2-\delta_{m,l})\;
\prod\limits_{\alpha=k}^{m-1}\left(
\frac{A^2}{\omega_{\alpha-1}\omega_{\alpha}}\right)^2
\prod\limits_{\beta=m}^{l-1}\left(
\frac{A^2}{\omega_{\beta-1}\omega_{\beta}}\right)\nonumber
\end {eqnarray}
In this relation, the term $(2-\delta_{m,l})$ is due to the binomial
coefficients in Eq.(\ref{aps3}). Following the result for $\omega_i$
and the functions $p_i$ and $q_i$ we can calculate the product:
\begin {equation}
\prod\limits_{\alpha=k}^{m-1}\left(
\frac{A^2}{\omega_{\alpha-1}\omega_{\alpha}}\right)=
\prod\limits_{\alpha=k}^{m-1}\left( \frac{A^2}
{\frac{p_{\alpha-1}}{q_{\alpha-1}}\frac{p_{\alpha}}{q_{\alpha}}}\right)
= \frac{Q_{k-1}Q_{k}}{Q_{m-1}Q_{m}}
\end {equation}
where we have introduced the more convenient variables:
$$ Q_{i}=\frac{q_i}{A^i} = w_1\left(\frac{\rho_1}{A}\right)^i +
w_2\left(\frac{\rho_2}{A}\right)^i$$ and also we performed the
replacement:
$$ \frac{A^2}{\omega_{k-1}^2} = \frac{Q_{k-1}^2}{Q_k^2} $$
Inserting the last results into Eq.(\ref{aps4}) we find:
\begin {equation}
\mathcal{S}_N(j_1+j_2+\cdots+j_N=1)=1-\frac{\pochh{1/2}{2}}{2A^2
z^2} \sum\limits_{k=1}^{N-1}\;\left[2\; \sum\limits_{m=k}^{N-1}\;
\sum\limits_{l=m}^{N-1}\;\frac{Q_{k-1}^4}{Q_{m-1}Q_m Q_{l-1}Q_l}-
\sum\limits_{l=k}^{N-1}\;\frac{Q_{k-1}^4}{ Q^2_{l-1}Q^2_l}\right]
\label{aps5}
\end {equation}
In the above equation we have the object symmetric in the indices
$m$ and $l$
$$S_{m,l}=\frac{1}{Q_{m-1}Q_m Q_{l-1}Q_l}$$ This symmetry significantly
simplifies  the calculation of the double summation, because we
can proceed as:
$$\sum\limits_{m=k}^{N-1}\;\sum\limits_{l=m}^{N-1}\;S_{m,l}\;=\;
1/2\left\{\sum\limits_{m=k}^{N-1}\;\sum\limits_{l=m}^{N-1}\;S_{m,l}
+\sum\limits_{l=k}^{N-1}\;\sum\limits_{m=l}^{N-1}\;S_{l,m}\right\}$$
In the second term we interchange the order of the summations by
identity:
$$\sum\limits_{l=k}^{N-1}\;\sum\limits_{m=l}^{N-1}\;S_{l,m} =
\sum\limits_{m=k}^{N-1}\;\sum\limits_{l=k}^{m}\;S_{l,m}$$ Finally,
we find:
\begin {equation}
\sum\limits_{m=k}^{N-1}\;\sum\limits_{l=m}^{N-1}S_{l,m}\;=\;
1/2\left\{\sum\limits_{m=k}^{N-1}\;\sum\limits_{l=k}^{N-1}\;S_{m,l}-
\sum\limits_{m=k}^{N-1}\;S_{m,m}\right\} \label{aps7}
\end {equation}
where the identity can be used:
$$\sum\limits_{m=k}^{N-1}\;\sum\limits_{l=k}^{N-1}\;S_{m,l}\;=\;
\left(\sum\limits_{l=k}^{N-1}\frac{1}{Q_{l-1}Q_l}\right)^2$$ To
calculate the sum, we introduce the relation:
$$\tilde{Q_i} = w_1x^i-w_2y^i\ .$$
Let us remind, that $$Q_i = w_1x^i+w_2y^i$$ where
$$x,y\;=\;\frac{\rho_{1,2}}{A}\; = \;
\frac{1}{2A}\pm\sqrt{\frac{1}{4A^2}-1}$$ Using the identity:
$$\frac{\tilde{Q}_l}{Q_l}\;-\;\frac{\tilde{Q}_{l-1}}{Q_{l-1}}\;=\;
\frac{2w_1w_2(x-y)}{Q_{l-1}Q_l}$$ we are able to calculate the
sum:
$$\sum\limits_{l=k}^{N-1}\frac{1}{Q_{l-1}Q_l}\;=\;
\frac{1}{2w_1w_2(x-y)}\left(\frac{\tilde{Q}_{N-1}}{Q_{N-1}}\;-\;
\frac{\tilde{Q}_{k-1}}{Q_{k-1}} \right)$$ Inserting these
intermediate results into Eq.(\ref{aps5}), we find:
\begin {equation}
\mathcal{S}_N(j_1+j_2+\cdots+j_N=1)=\;
 1-\frac{\pochh{1/2}{2}}{2A^2 z^2}\;\sum\limits_{k=1}^{N-1}\
Q_{k-1}^4\;
\left[\left(\sum\limits_{l=k}^{N-1}\frac{1}{Q_{l-1}Q_l}\right)^2-
2\sum\limits_{l=k}^{N-1}\frac{1}{Q^2_{l-1}Q^2_l}\right]
\label{vysl2}
\end {equation}
 The sum over $k$ is the problem of
the sum of the finite power series. For the finite $N$ we found
the result:
\begin {eqnarray}
\mathcal{S}_N&(&j_1+j_2+\cdots+j_N=1)=\; 1 -
\frac{\pochh{1/2}{2}}{2A^2z^2}\;\frac{1}{(x-y)^3}
\label{aps6}\\
\noalign{\vskip8pt}
 &\times&\left\{\frac{x^{2N-2}-y^{2N-2}}{(x+y)}
\left[\frac{1}{x^2(x^{N-1}+\frac{w_2}{w_1}y^{N-1})^2} +
\frac{1}{y^2(y^{N-1}+\frac{w_1}{w_2}x^{N-1})^2}\right]\right.\nonumber\\
\noalign{\vskip8pt} &+&
2\left[\frac{(1-x^{2N-2})(1-\frac{w_2}{w_1}y^{2N-2})}
{x(x^{N-1}+\frac{w_2}{w_1}y^{N-1})^2} -
\frac{(1-y^{2N-2})(1-\frac{w_1}{w_2}x^{2N-2})}
{y(y^{N-1}+\frac{w_1}{w_2}x^{N-1})^2}\right]\nonumber\\
\noalign{\vskip8pt}
 &+& \left.\frac{(x-y)(N-1)}{2} \left[3\;
\frac{(x^{N-1}-\frac{w_2}{w_1}y^{N-1})^2}
{(x^{N-1}+\frac{w_2}{w_1}y^{N-1})^2}\;-\; 1
\right]\right\}\nonumber
\end {eqnarray}

In the above relation we have all the symbols defined in the
previous text. The continuum limit means to take the limit
$N\rightarrow \infty$ in all  formulas obtained for finite $N$.
Doing this, we find for symbols appeared in (\ref{aps6}):
\begin {eqnarray*}
\frac{w_2}{w_1}& \rightarrow & 1\\
x-y & \rightarrow & 2\triangle\gamma\\
x,y & \rightarrow & 1\\
x^N & \rightarrow & e^{\beta\gamma}\\
y^N & \rightarrow & e^{-\beta\gamma}\\
\gamma & \rightarrow & \sqrt{\frac{2b}{c}}\\
z^{-2} & \rightarrow & 2a\triangle^3/c^2\\
\triangle & \rightarrow & \frac{\beta}{N}
\end {eqnarray*}
Inserting into the Eq.(\ref{aps6}) we find finally for the first
nontrivial term in the continuum limit the result:
\begin {equation}
\mathcal{S}_N(j_1+j_2+\cdots+j_N=1)=\; 1
-\frac{\pochh{1/2}{2}}{2}\; \frac{a}{c^2\gamma^3}\;
\left\{\frac{e^{2\beta\gamma} - e^{-2\beta\gamma}}
{(e^{\beta\gamma}+ e^{-\beta\gamma})^2} + \beta\gamma\left[
3\frac{(e^{\beta\gamma}- e^{-\beta\gamma})^2} {(e^{\beta\gamma}+
e^{-\beta\gamma})^2}-1\right]\right\}
\end {equation}
In derivation of the above relation no connection between the
variables $a$ and $b$ has been supposed. The relation is finite in
the whole range of the values of $b$, positive, zero or negative.
For large $b$ and moderate $a$ our result corresponds to the
results of the conventional perturbative method of calculation.

\section*{Appendix 3: The evaluation of the functional integral by the Gelfand-Yaglom procedure}
\appendix

The well-known method of the evaluation of the functional integral
for the harmonic oscillator in the continuum limit was proposed by
Gelfand-Yaglom \cite{bkz}. We apply the same idea for the
evaluation of the functional integral with the fourth order term
in the potential.

In the $N-th$ approximation for such functional integral we find
in the Appendix 1 the result (\ref{vysl1}) and in the Appendix 2
the approximation of this result (\ref{vysl3}),(\ref{vysl2}). The
common form of the both relations for $N-th$ approximation can be
expressed as: $$\mathcal{Z}_N =
\frac{S_{N-1}(\triangle)}{\sqrt{\prod\limits^{N-1}_{i=0}2(1+b\triangle^2/c)\omega_i}}$$
The value of the functional integral in the continuum limit is
defined as $$\mathcal{Z} = \lim_{N \rightarrow
\infty}\mathcal{Z}_N$$

Let us define the function
\begin {equation}
F_k =
\frac{\prod\limits^{k}_{i=0}2(1+b\triangle^2/c)\omega_i}{S^2_{k}(\triangle)}
\label{gjf}
\end {equation}
The function $S_{k}(\triangle)$ is the result after first $k$
steps od the calculation of the function $S_{N}(\triangle)$, as it
was demonstrated in the Appendices 1 and 2. Let us stress that
$$\mathcal{Z}_N = \frac{1}{\sqrt{F_N}}$$
The aim of the Gelfand-Yaglom construction is to find in the
continuum limit the differential equation, such that its solution
is connected to the continuum functional integral by:
$$\mathcal{Z} =\frac{1}{\sqrt{F(\beta)}}.$$

In the sense of the Gelfand-Yaglom construction we are going to
express the relation for $F_{k+1}$ by help of $F_{k}$ and $F_{k-1}$.
We have:
\begin {equation}
F_{k+1}=\frac{2(1+b\triangle^2/c)\prod\limits^{k}_{i=0}2(1+b\triangle^2/c)\omega_i}{S^2_{k+1}(\triangle)}-
\frac{\prod\limits^{k-1}_{i=0}2(1+b\triangle^2/c)\omega_i}{S^2_{k+1}(\triangle)}
\label{gyeq3}
\end {equation}
To obtain the above equation we used (\ref{gjf}) and the identities:
$$\omega_i=1-A^2/\omega_{i-1},\; \omega_0=1/2+b \triangle^2/c,\;
A=\frac{1}{2(1+b\triangle^2/c)\; .}$$

To introduce $F_{k}$ and $F_{k-1}$, we must to express
$S^2_{k+1}(\triangle)$ by help of $S^2_{k}(\triangle)$ and
$S^2_{k-1}(\triangle)$ in the corresponding terms. This means, that
we use:
\begin {eqnarray}
S_{k+1}(\triangle) &=& S_{k}(\triangle)+(S_{k+1}(\triangle)-S_{k}(\triangle))\\
S_{k+1}(\triangle) &=& S_{k-1}(\triangle)+
(S_{k+1}(\triangle)-S_{k-1}(\triangle))\nonumber
\end {eqnarray}
In the case of the calculation as in the Appendix 2 the functions
$S_{k}(\triangle)$ can be calculated without approximations. The
explicit calculations shown, that
$(S_{k+1}(\triangle)-S_{k}(\triangle))/S_{k}(\triangle)\approx
\triangle$ also for the non-perturbative anzatz of the Appendix 1.
Therefore we expands the denominators in (\ref{gyeq3}) up to the
terms of the second power in $\triangle$ and we obtain:
\begin {eqnarray}
\frac{1}{S_{k+1}(\triangle)^2} &=&
\frac{1}{S_{k}(\triangle)^2}\left[1-
2\frac{S_{k+1}(\triangle)-S_{k}(\triangle)}{S_{k}(\triangle)}
+3\left(\frac{S_{k+1}(\triangle)-S_{k}(\triangle)}{S_{k}(\triangle)}\right)^2+...
\right]\\
\frac{1}{S_{k+1}(\triangle)^2} &=&
\frac{1}{S_{k-1}(\triangle)^2}\left[1-
2\frac{S_{k+1}(\triangle)-S_{k-1}(\triangle)}{S_{k-1}(\triangle)}
+3\left(\frac{S_{k+1}(\triangle)-S_{k-1}(\triangle)}{S_{k-1}(\triangle)}\right)^2+...
\right]\nonumber
\end {eqnarray}
After some algebra we we find for the Eq.(\ref{gyeq3}) the
difference equation:
\begin {equation}
\frac{F_{k+1}-2F_{k}+F_{k-1}}{\triangle^2}+
4\frac{F_{k}-F_{k-1}}{\triangle}\frac{S_{k+1}-S_{k}}{\triangle\;
S_{k}} = F_k\left[2b/c
-2\frac{S_{k+1}-2S_{k}+S_{k-1}}{\triangle^2\;
S_{k}}-2\left(\frac{S_{k+1}-S_{k}}{\triangle\;
S_{k}}\right)^2\right]
\end {equation}
In the continuum limit $\triangle \rightarrow 0$ we use the notation
$$k \triangle \rightarrow \tau$$ and we finally find the
differential equation:
\begin {equation}
\frac{\partial^2}{\partial
\tau^2}F(\tau)+4\frac{\partial}{\partial \tau}F(\tau)\,
\frac{\partial}{\partial \tau}\ln{S(\tau)}=
F(\tau)\left(\frac{2b}{c}-2\frac{\partial^2}{\partial
\tau^2}\ln{S(\tau)}-4(\frac{\partial}{\partial \tau}\ln{S(\tau)})^2\right)
\end {equation}
The initial conditions are:
\begin {eqnarray}
F(0) &=& 1,
\label{gyeqa2} \\
\frac{\partial}{\partial \tau}F(0) &=& 0.\nonumber
\end{eqnarray}
By substitution $$F(\tau) = \frac{y(\tau)}{S^2(\tau)}$$ the
differential equation can be read in more convenient form:
\begin {equation}
\frac{\partial^2}{\partial \tau^2}y(\tau) =\frac{2b}{c}\  y(\tau)
\end {equation}
The importance of the thorough evaluation of the function $S(\tau)$ emerged
in this construction of generalized Gelfand-Yaglom equation.

\section*{Appendix 4: Conditional Wiener measure}
\appendix

 The conditional measure functional integrals define the
propagators in the theory, contrary to the unconditional measure
functional integrals, which define the partition function. The
 conditional Wiener measure functional integral is defined as the
continuous limit of the finite dimensional integral with fixed
endpoints:

\begin {equation} I=\lim_{N\rightarrow\infty}\int\limits
_{-\infty}^{+\infty} \prod \limits
_{i=1}^{N-1}\left(\frac{d\varphi_i}{\sqrt{\frac{2\pi\triangle}{c}}}\right)
\exp\left\{-\sum\limits _{i=1}^N \triangle\left[c/2
\left(\frac{\varphi_i-\varphi_{i-1}}{\triangle}\right)^2
+b\varphi_i^2+a\varphi_i^4\right]\right\} \label {eqcon1}
\end {equation}

\noindent The only difference between the unconditional and the
conditional definition inhere in the dimension of the finite
dimensional integral, whereas the actions are the same. For
conditional case, both the endpoints of $\varphi$ in the time
variable are fixed. This correspond to calculation of the
propagator of the model, whereas the unconditional measure
integral correspond to calculations of the partition functions.
The result of the finite dimensional integration in Eq.(\ref
{eqcon1}), up to index $k_{N-2}$ is the same as for the
conditional measure case. For the last time slice integration, for
$k_{N-1}$ we have:

\begin {equation}
\sum\limits_{k_{N-1}=0}^{\infty}\;
\frac{(\frac{c}{\triangle}\varphi_N)^{2k_{N-1}}}{(2k_{N-1})!}\;
\int\limits_{-\infty}^{\infty}
\frac{d\varphi_{N-1}}{\sqrt{\frac{2\pi\triangle}{c}}}\
(\varphi_{N-1})^{2k_{N-2}+2k_{N-1}} \;
\exp{-\{a\triangle\varphi_{N-1}^4+c/\triangle(1+b\triangle^2/c)\varphi_{N-1}^2\}}
\end {equation}

\noindent The conditional measure result we derive from the result
for the unconditional case, when the term depending on the
summation index $k_{N-1}$ we replace by the term:

\begin {equation}
\sum\limits_{k_{N-1}=0}^{\infty}\;
\frac{(\frac{c\varphi_N^2}{\triangle(1+b\triangle/c)})^{k_{N-1}}}{(2k_{N-1})!}\;
\Gamma(k_{N-2}+k_{N-1}+1/2)\mathcal{D}_{-k_{N-2}-k_{N-1}-1/2}\;
(z)\; e^{-\{a\triangle\varphi_N^4\ + \
\frac{c}{\triangle}(1/2+\frac{b\triangle^2}{c})\varphi_N^2\}}
\end {equation}

In this appendix we do not follow by discussion for the arbitrary
value of the $\varphi_N$, we are going to discuss the simplest
case of the $\varphi_N=0$, corresponding to the periodic boundary
conditions in the imaginary time variable. The result of the $N-1$
dimensional integration can be read:

\begin {equation}
\sum \limits_{k_1,\cdots,k_{N-1}}^\infty \prod \limits
_{i=1}^{N-1} \; \left[\frac{1}{\sqrt{2\pi(1+b\triangle/c)}} \;
\frac{(\xi/z)^{2k_{i}}}{(2k_{i})!}
\Gamma(k_{i-1}+k_{i}+1/2)\mathcal{D}_{-k_{i-1}-k_{i}-1/2}\;(z)\right]
e^{-\{a\triangle\varphi_N^4\ + \
\frac{c}{\triangle}(1/2+\frac{b\triangle^2}{c})\varphi_N^2\}}
\label{con4}
\end {equation}

\noindent In the last equation the identity $k_0=0$ is required. The
main difference from the unconditional measure calculation consists
in the fact, that the arguments of the parabolic cylinder functions
in the Eq.(\ref{con4}) are the same in contrary to unconditional
measure case, where the argument of the last function differs from
the all others arguments of the $\mathcal{D}$ functions. We are
going to repeat  the calculation as in the Appendix 2, to show the
differences of the unconditional measure calculations from
conditional measure ones  for the first two terms of the
decomposition of the corresponding functional integrals. We again
introduce the objects as for unconditional measure case, but
corresponding to the conditional measure specifications:
$$\omega_i=1-\frac{A^2}{\omega_{i-1}}$$ with
$$\omega_0=1\ .$$  Following the
unconditional measure calculation, by the n-th convergent method
we find: If $\omega_n$ is the continued fraction defined by:
$$\omega_n = a_1+\frac{b_1}{a_2+\frac{b_2}{a_3+\cdots}}$$ then we
can find the $\omega_n$ in the form of the $n-th$ convergent:
$$\omega_n = \frac{p_n}{q_n}$$ where $p_n$ and $q_n$ are the solutions
of the homogeneous difference equations with the constant
coefficients:

\begin {eqnarray}
p_n &=& a_n \;p_{n-1}+b_n \; p_{n-2}\nonumber\\
q_n &=& a_n \;q_{n-1}+b_n \; q_{n-2}\nonumber\\
\end {eqnarray}

\noindent The solutions have the form:
\begin {eqnarray}
p_n &=& \tilde{w}_1 \rho_1^n + \tilde{w}_2 \rho_2^n\nonumber\\
q_n &=& w_1 \rho_1^n + w_2 \rho_2^n \nonumber\\
\end {eqnarray}

\noindent where $\rho_{1,2}$ are the solutions of the
characteristic equation:
$$ \rho^2 -a_n\rho-b_n = 0 $$
where
 $$a_n=1$$
 $$b_n=-A^2.$$
 \noindent
The boundary conditions for $p_n,\; q_n$ are fixed by the first
two relations for $\omega_n$:
$$\omega_0=1$$
$$\omega_1=1-A^2$$
and we have:
\begin {eqnarray}
p_0 &=& 1\\ p_1 &=& 1-A^2\\ q_0 &=& 1\\ q_1 &=& 1
\end {eqnarray}
\noindent The $n-th$ convergent solution for $\omega_n$ is thus
given by the relations:
\begin {eqnarray}
\rho_{1,2} &=& \frac{1}{2}(1\pm \sqrt{1-4A^2}\;)\\
\tilde{w}_{1,2} &=& \frac{1}{2}\left(1\pm
\frac{1-2A^2}{\sqrt{1-4A^2}}\right)\\
w_{1,2} &=&\frac{1}{2}\left( 1 \pm \frac{1}{\sqrt{1-4A^2}}\right)
\end {eqnarray}

\noindent We see from the last relations the very powerful
characteristic of the solution represented by the identity:
$$p_n=q_{n+1},\ n=1,2,\cdots$$

Expanding the $N-1$ dimensional integral for the conditional
measure into the asymptotic series, we find the relation:
\begin {equation}
\mathcal{S}_N
=\left\{\prod\limits_{i=1}^{N-1}[2(1+b\triangle^2/c)\omega_i]\right\}^{-1/2}\
\mathcal{Z}_N
\end {equation}

The first part of this relation in the continuum limit is the
conditional measure functional integral determinant for the
harmonic oscillator type action. The second part, marked
$\mathcal{Z}_N$, is the asymptotic expansion correction. As for
the unconditional measure integral we calculated the first two
corrections terms and we find:
\begin {eqnarray}
1 &-& \frac{\pochh{\frac{1}{2}}{2}}{2z^2}\;\frac{1}{(x-y)^3}\\
\noalign{\vskip8pt}
 &\times&\left\{\frac{x^{2N-2}-y^{2N-2}}{(x+y)}
\left[\frac{1}{x^2(x^{N-1}+\frac{w_2}{w_1}y^{N-1})^2} +
\frac{1}{y^2(y^{N-1}+\frac{w_1}{w_2}x^{N-1})^2}\right]\right.\nonumber\\
\noalign{\vskip8pt} &+&
2\left[\frac{(1+\frac{w_2}{w_1}-x^{2N-2}-\frac{w_2}{w_1}y^{2N-2})}
{x(x^{N-1}+\frac{w_2}{w_1}y^{N-1})^2} -
\frac{(1+\frac{w_1}{w_2}-y^{2N-2}-\frac{w_1}{w_2}x^{2N-2})}
{y(y^{N-1}+\frac{w_1}{w_2}x^{N-1})^2}\right]\nonumber\\
\noalign{\vskip8pt}
 &+& \left.\frac{(x-y)(N-1)}{2} \left[3\;
\frac{(x^{N-1}-\frac{w_2}{w_1}y^{N-1})^2}
{(x^{N-1}+\frac{w_2}{w_1}y^{N-1})^2}\;-\; 1
\right]\right\}\nonumber
\end {eqnarray}

Above relation possesses the well defined continuum limit. We
find, in the limit $N\rightarrow\infty$ the following relations:
\begin {eqnarray*}
\frac{w_2}{w_1}& \rightarrow & -1\\
x-y & \rightarrow & 2\triangle\gamma\\
x,y & \rightarrow & 1\\
x^N & \rightarrow & e^{\beta\gamma}\\
y^N & \rightarrow & e^{-\beta\gamma}\\
\gamma & \rightarrow & \sqrt{\frac{2b}{c}}\\
z^{-2} & \rightarrow & 2a\triangle^3/c^2\\
\triangle & \rightarrow & \frac{\beta}{N}
\end {eqnarray*}
\noindent then, finally, in the continuum limit the first nontrivial
correction to $\mathcal{Z}_N$ is:
\begin {equation}
1-\frac{\pochh{\frac{1}{2}}{2}a}{c^2}\;
\frac{1}{8\gamma^3\left(e^{\gamma\beta}-e^{-\gamma\beta}\right)^2}
\left\{-3(e^{2\gamma\beta}-e^{-2\gamma\beta})+
2\gamma\beta\left[e^{2\gamma\beta}+e^{-2\gamma\beta}+4\right]\right\}
\end {equation}
or,
\begin {equation}
1- \frac{3 a}{32c^2\gamma^3} \left\{-3\coth(\gamma\beta)+
2\gamma\beta\left[\coth^2(\gamma\beta)+\frac{1}{2\sinh^2(\gamma\beta)}\right]\right\}
\end {equation}
We can stress, that our result is valid for finite values of the
parameter "$b$" positive or negative, so we describe the case of the
anharmonic oscillator and the Higgs model as well.

\end {document}